\documentclass[12pt]{iopart}
\usepackage{graphicx}
\newcommand{\nn}{\nonumber}

\newcommand{\eq}[1]{
\begin{equation}
#1
\end{equation}} % short form of the equation environment

\newcommand{\eqq}[2][]{
\begin{equation}
#2 \label{#1}
\end{equation}} % same as before but with a label argument defined right after the command

\newcommand{\ee}[1]{
\begin{eqnarray}
#1
\end{eqnarray}}

\newcommand{\eem}[2][]{
\begin{eqnarray}
#2 \label{#1}
\end{eqnarray}} 
\begin{document}

\title
{Inhomogeneous Quantum Quenches}

\author
{Spyros Sotiriadis$^{1}$ and John Cardy$^{1,2}$ }

\address{$^{1}$ Oxford University, Rudolf Peierls Centre for Theoretical Physics, 1 Keble
Road, Oxford OX1 3NP, UK } 
\address{$^{2}$ All Souls College, Oxford, UK}
\eads{\mailto{s.sotiriadis1@physics.ox.ac.uk}, \mailto{j.cardy1@physics.ox.ac.uk}}

\date{\today}

\begin{abstract}
We study the problem of a quantum quench in which the initial state is the ground state of an inhomogeneous hamiltonian, in two different models, conformal field theory and ordinary free field theory, which are known to exhibit thermalisation of finite regions in the homogeneous case. We derive general expressions for the evolution of the energy flow and correlation functions, as well as the entanglement entropy in the conformal case. Comparison of the results of the two approaches in the regime of their common validity shows agreement up to a point further discussed. Unlike the thermal analogue, the evolution in our problem is non-diffusive and can be physically interpreted using an intuitive picture of quasiparticles emitted from the initial time hypersurface and propagating semiclassically.
\end{abstract}


\maketitle

\section{Introduction}
%\emph{Definition of an inhomogeneous quantum quench. Discussion of earlier work. 
%Description of main results. Outline of paper.}

An isolated quantum system in which some of the parameters that determine the dynamics change rapidly at a specific instant is said to undergo a quantum quench. Recently such rapid changes have become experimentally feasible in cold atom systems \cite{ex:cr-02, ex:kww-06,ex:spin-06,ex:tu-06}. These experiments as well as the development of numerical techniques for the study of quantum dynamics (t-DMRG) have motivated significant theoretical interest as the investigation of a wide range of problems, either in lattice models \cite{sps-04,dmcf-06,cc-06,cc-07,ep-07,l-07,kol-07,qr-08,kol-08,fc-08} or in continuous theories \cite{cc-06,cc-07,cc-07b} shows. The importance of quantum quenches relies on the potential discovery of novel physical phenomena and the fact that they will help us to better understand out of equilibrium quantum behaviour. On the other hand they can be described by simple theoretical models. More specifically, to find the evolution that follows a quantum quench in the Heisenberg picture one should calculate the expectation values of operators whose time dependence is determined by the hamiltonian after the quench, with respect to the pure state in which the system was before the quench, which is typically the ground state of the initial hamiltonian. Then one finds that while in systems with finite number of degrees of freedom local observables exhibit periodic or quasiperiodic behaviour, in the thermodynamic limit they tend to stationary values instead. It should be emphasised that this happens to \emph{connected} correlation functions of \emph{local} observables, as a result of the interference of the infinite number of momentum modes of the whole system. 

In particular let us consider a system of coupled harmonic oscillators or equivalently a \emph{free} field theory, described by a general dispersion relation with some energy gap or ``mass" $m_0$ and maximum group velocity of excitations $v_{max}$. Assume that the system lies on the ground state of the initial hamiltonian $H_0$ when at time $t=0$ the mass is quenched from $m_0$ to a different value $m \neq 0$. Then after the quench there is an extensive excess in energy in comparison with the ground state of the final hamiltonian $H$, which is distributed to the excitation levels of $H$. In a spacetime representation these excitations appear as quasiparticles that emerge from the $t=0$ hypersurface and propagate forward in time. For free field theories the two-point correlation function contains all the information required to determine their state since this is a superposition of gaussian wavepackets. This correlation function of two points separated by distance $r$ turns out \cite{cc-06,cc-07} to remain  unaffected by the quench until time $t=r/2 v_{max}$ when it starts changing and finally for large times it takes the form of a thermal correlation function with a \emph{momentum dependent} effective temperature. Additionally if $m_0>m$ then for large values of $m_0$ the effective temperature $1/\beta_{eff}$ is of order $m_0$ and asymptotically independent of the momentum \cite{talk}. To understand this behaviour we can imagine pairs of coherent quasiparticles emitted by neighbouring %(and so initially correlated) 
points of the $t=0$ hypersurface that induce correlations between spatially separated points as soon as the fastest ones reach them after time $t=r/2 v_{max}$. This is called the \emph{horizon effect} \cite{cc-06,cc-07}. On the other hand for large times, the interference between a large number of incoherent quasiparticles coming from different and uncorrelated points of the $t=0$  hypersurface drives the system to stationary and in particular thermal behaviour as far as \emph{local} observables are concerned. This means that any finite subsystem tends to thermal equilibrium with its complement which acts as a bath (\emph{thermalisation}) \cite{rig-06,rig-07,cc-06,cc-07,qr-08} (also \cite{rig-08,bs-08} for recent general discussions of the subject). The fact that the effective temperature depends on the momentum $k$ in the free field theory case should be expected since the final hamiltonian can be diagonalised exactly in momentum space which means that the different momentum modes evolve independently and so thermalise to different temperatures. 

The above observations have been shown to be valid not only in the simple case of a free field theory but also in two other important general cases. First \cite{cc-06,cc-07} in conformal field theory (CFT), which describes one-dimensional quantum systems at criticality in the continuum limit (equivalently massless \emph{interacting} field theories). Second \cite{sc-} %add
in a self-consistent Hartree-Fock approximation of a system of \emph{anharmonic} oscillators (equivalently an \emph{interacting} bosonic field theory), which is valid in the large-$N$ limit of the linear sigma model. 

An important question is to what extent the stationary behaviour of the system depends on the initial state and if there are any physical quantities other than the conserved energy (and possibly any other quantity that commutes with the hamiltonian) about which information survives in the final state. One way to study the effect of different initial states on the evolution is to choose the hamiltonian before the quench to be spatially inhomogeneous. We call this kind of quench an \emph{inhomogeneous} quench. More specifically we will concentrate on the previous system of harmonic oscillators where the initial mass $m_0$ now depends on the position $x$. In this case the initial hamiltonian is not diagonal in momentum space although it is still diagonalizable. A question of particular interest is whether the evolution after the quench resembles that of the thermal analogue, which would be the heat diffusion occurring in a system with inhomogeneous initial temperature distribution $u_0(x)$. A special case of distribution where this comparison should be easier is the step distribution, that is when $u_0(x)$, respectively $m_0(x)$, has different values for $x>0$ and $x<0$. In the thermal case the diffusive nature of the heat equation leads to the heat current across $x=0$ decreasing with time as $1/\sqrt{t}$ and we wish to study what happens following an equivalent initial state in the case of a quantum quench. To this end we will be calculating the energy flow instead, since there is not such a notion as the heat current in our case. Recall that in the thermal problem the heat current $j_{q}$ is related to the energy current $j_{e}$ and the particle current $j_{n}$ according to $j_{q} = j_{e} - \mu j_{n}$ where $\mu$ is the chemical potential, and that these currents also exhibit the same diffusive behaviour. 

At the same time we will derive general expressions for the correlation functions and the entanglement entropy which measure the correlations between points or parts of the system respectively. Similar calculations in the special case of a step distribution and $m=0$ have been done using a CFT method alternative to ours in \cite{chd-08}. Also local effects caused by defects have been investigated in \cite{ep-07} for a general lattice model and in \cite{cc-07b} for a continuous field theory. General analytical results for the calculation of the entanglement entropy in field theory can be found in \cite{hlw-94,ch-04,cc-04,ch-07}.

In this paper we consider a continuous bosonic field with a relativistic dispersion relation $\omega = \sqrt{k^2+m^2}$ where $v_{max}=1$, although other dispersion relations drawn from lattice models and having the same energy gap and maximum group velocity are expected to lead to similar behaviour. For our purposes we can choose the initial distribution to depend only on one space coordinate and so it is sufficient to consider only one dimensional systems. Also we often make use of the so-called \emph{deep quench} limit, that is the limit when $m_0 \gg m$. This must obviously reflect all the characteristic features of a quantum quench since it is one of the two most extreme possibilities for the relation between the two masses. In our inhomogeneous problem this means that $m$ should be much smaller than any value of the initial mass distribution $m_{0}(x)$.

This paper is organised as follows: in section \ref{sec:2} we apply the methods of CFT to solve the general problem in the case where the theory after the quench is massless. In section \ref{sec:3} we analyse two special cases where the initial distribution has a bump or a step. % and . %change this
In section \ref{sec:4} we use free field theory methods to solve the massive problem as well and in section \ref{sec:5} we compare the results obtained from these two approaches. Lastly in section \ref{sec:6} we discuss our findings as compared to the thermal analogue and give a physical interpretation.

\section{Massless case - CFT approach \label{sec:2}}
%\emph{Description of the geometry.  Conformal mapping to the constant width strip and derivation of $F(x)$.}

As shown in earlier work \cite{cc-07}, the problem of a quantum quench can be mapped to a Euclidean field theory defined on a strip (or a $d$+1-dimensional slab in general) where the transverse direction corresponds to imaginary time $\tau$. To see this let us consider the expectation value of a local operator after the quench
\eq{\langle\mathcal{O} (t,\{r_i\})\rangle = \langle\Psi_0|e^{iHt}\mathcal{O} (\{r_i\})e^{-iHt}|\Psi_0\rangle}
where $|\Psi_0\rangle$ is the initial state and $H$ is the hamiltonian after the quench. The last relation can be written in path integral form and to assure its convergence one should first insert damping factors $e^{-\epsilon H}$ with $\epsilon\to 0$ as follows
\eqq[obs]{\langle\mathcal{O} (t,\{r_i\})\rangle = Z^{-1} \langle\Psi_0|e^{iHt-\epsilon H}\mathcal{O} (\{r_i\})e^{-iHt-\epsilon H}|\Psi_0\rangle}
where $Z=\langle\Psi_0|e^{-2\epsilon H}|\Psi_0\rangle$ is a normalisation factor. If we analytically continue to imaginary time $\tau$ then we obtain the same expression as that corresponding to a strip of width $2\epsilon$ where $|\Psi_0\rangle$ plays now the role of boundary conditions along both borders of the strip. Using the arguments of Renormalisation Group (RG) theory one can argue that as long as $|\Psi_0\rangle$ is translationally invariant, it can be safely replaced by another state that corresponds to an RG-invariant boundary condition, without changing the asymptotic behaviour of (\ref{obs}) in the limit $\epsilon\to 0$. In particular it turns out that for the bosonic field we discussed in the introduction and the deep quench limit, we have to impose Dirichlet boundary conditions (b.c.) forcing the field to vanish on the boundary. The same RG arguments give $\epsilon$ a physical meaning as the typical time scale of the dynamics near the ground state of $H_0$, that is the inverse initial mass $1/m_0$. 

\begin{figure}[htbp]
\begin{center}
\includegraphics[width=1.00\textwidth]{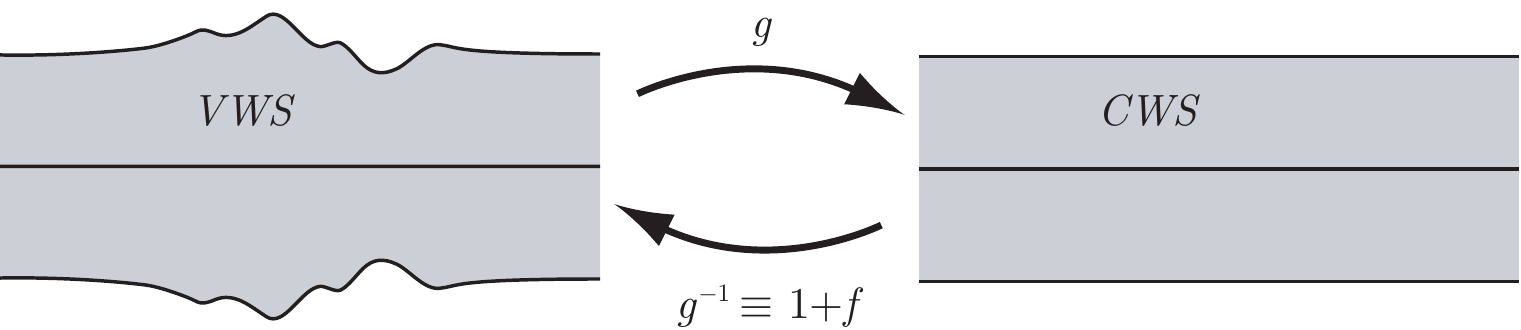}
\caption{Schematic representation of the transformation from the VWS to the CWS.}
\label{fig:map}
\end{center}
\end{figure}

This suggests that an inhomogeneous quench can still be formulated on a strip with Dirichlet b.c. but having variable width $2 (\epsilon + h(x))$ where $h(x)$ expresses the variation of the initial mass $2 (\epsilon + h(x)) \sim m_{0}^{-1}(x)$. Now let us assume that the hamiltonian after the quench is massless, i.e. $m=0$. In this case we can use CFT techniques and solve the problem directly for general $h(x)$ by exploiting the conformal invariance of the theory. Indeed if we map the variable width strip (VWS) to the simpler geometry of a strip with constant width $2 \epsilon$ (CWS) using a conformal mapping $w \to z=g(w)$ then the transformation law of correlation functions of (primary) operators under such mappings will allow us to derive the corresponding expressions in the VWS from those in the CWS which are already known (Fig.\ref{fig:map}). 
The appropriate conformal mapping should satisfy the condition that 
\eqq[bc1]{\mbox{Im}g(x\pm i(\epsilon+h(x)))=\pm \epsilon }
so that it deforms the boundaries as required. Notice that unlike the convention used in previous papers that the strip ranges from $\tau=0$ to $\tau=2\epsilon$, we now center the strip in the middle so that $\tau$ ranges from $-\epsilon$ to $+\epsilon$. This is a well-posed mathematical problem whose solution can be found  in general by solving Laplace's equation. In this way we obtain $g(w)$ as an integral along the boundary of the VWS, a form that depends implicitly on $h(x)$. This form is both cumbersome and actually inappropriate for our purposes since in order to extract useful information about physical quantities we only need to consider special asymptotic limits. To this end we will use two approaches: 

\begin{itemize}

\item the limit in which the transformation is infinitesimal, i.e. $m_{0}(x)$ varies only slightly in comparison with some average value. In this limit, if we define the inverse transformation as $1+f \equiv g^{-1}$, the boundary conditions for $f$ are to first order in $h$ 
\eqq[bcf]{\mbox{Im}f(x\pm i\epsilon)=\pm h(x)}
The solution can be found by using the general solution to the Laplace equation or by manipulating the analyticity of $f$. We then find that, up to an irrelevant real additive constant, $f$ is given by
\eqq[transf]{ f(z)  = \int\limits_{-\infty}^{+\infty}{ds \; h(s) F(z-s)} }
with kernel
\eqq[kernel]{F(z)= \frac{1}{\epsilon (e^{-\pi z/\epsilon}+1)} = \frac{1}{2 \epsilon}\left( \tanh{ \frac{\pi z}{2\epsilon}} + 1 \right)}
where we have assumed without loss of generality that $h(-\infty)=0$. In \ref{App1} we explain all of the above in detail and present an elegant derivation of the solution.

As we will see, to calculate the physical quantities we want it is sufficient to know the form of $f(z)$ for real values of $z$. In addition if the distribution $h(x)$ changes only slowly in comparison with $\epsilon$ then, assuming that the following expression is convergent, the asymptotic form of $f(x)$ in the limit $\epsilon \to 0$ that we are interested in, is
\eqq[transf2]{ f(x)  = \frac{1}{\epsilon}\int\limits_{-\infty}^{x}{ds \; h(s) } }
since $F(x-s)$ can be written in this limit as $\Theta(x-s)/\epsilon$.
\item 
the asymptotic behaviour of the transformation for large $|z|$, which turns out to determine the behaviour of physical quantities for large times and separations. 
Since the initial distribution $h(x)$ must be bounded, we can distinguish two important cases: either it tends to the same value as $x\to \pm \infty$ or to different values for each limit. In what follows we will consider one characteristic example for each case, focusing on the second which is more interesting: a \emph{bump} distribution and a \emph{step} distribution, both localised at the origin. Having assumed that $h(-\infty)=0$, for $\mbox{Re}z\to -\infty$ the transformation becomes asymptotically equal to the identity, that is $f(z) \to 0$. If we call $h(+\infty)=\alpha$ then for $\mbox{Re}z\to +\infty$ the transformation should rescale the strip width by $(1+\alpha/\epsilon)$ and since it is conformal must also rescale the $x$-direction by the same amount (Fig. \ref{fig:map2}). 
\begin{figure}[htbp]
\begin{center}
\includegraphics[width=.90\textwidth]{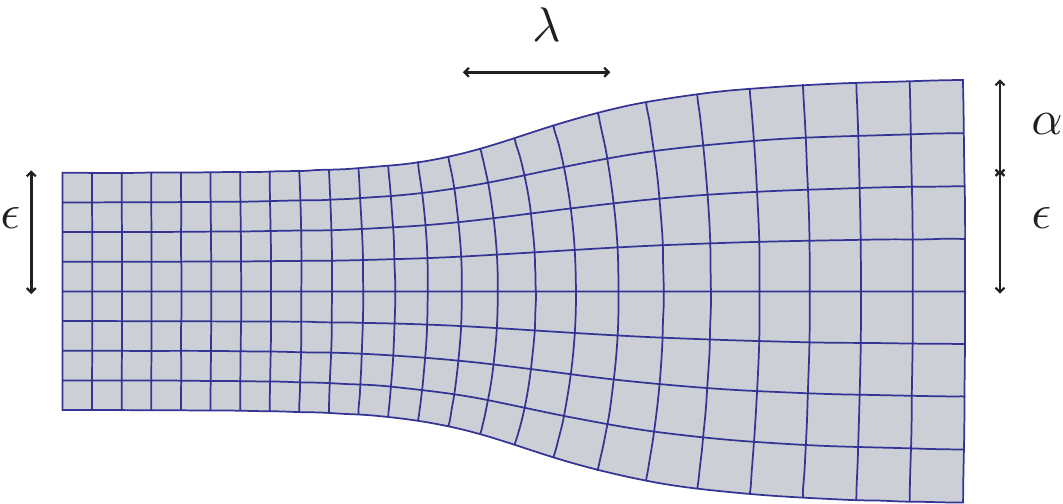}
\caption{Conformal map from the CWS to a stepped width strip, as illustrated by the deformation of gridlines of the CWS. The transformation used is $f(z) = \alpha \lambda \log\left( 1+ e^{2 z/\lambda} \right)/2\epsilon$ where $\alpha$ controls the increase in the width in the right half and $\lambda$ the distance over which the change happens. Notice that due to the fact that the transformation is conformal, it scales the strip in the longitudinal direction by the same amount as in the transverse direction, in the right half.}
\label{fig:map2}
\end{center}
\end{figure}
This means that the asymptotic form of $f(z)$ as $\mbox{Re}z\to + \infty$ can only be 
\eq{f(z) \sim \frac{\alpha}{\epsilon}z+\beta}
where $\beta$ is real so that, if $\alpha =0$, $f(z)$ corresponds to a translation along the $x$-direction for $\mbox{Re}z\to +\infty$. Overall the asymptotic form of the transformation from the CWS to the VWS is 
\eqq[asym]{z+f(z)\sim 
 \cases{  
  z  & \mbox{for Re$(z)\to-\infty$}  \\
  (1+\alpha /\epsilon) z + \beta & \mbox{for Re$(z)\to+\infty$}  \\
 } }
and that of the inverse transformation $g$
\eqq[gasym]{g(w)\sim
 \cases{  
  w  & \mbox{for Re$(w)\to-\infty$} \\
{(w-\beta)}/(1+\alpha/\epsilon)  & \mbox{for Re$(w)\to+\infty$}
 } }
The bump distribution corresponds to $\alpha =0$, while the step distribution to $\alpha \neq 0$ where $\alpha$ is the height of the step. 
It can be verified that the infinitesimal form of the transformation obtained from (\ref{transf}) and (\ref{kernel}) has the expected asymptotic behaviour in these two cases. Notice that due to the fact that $h(x)$ is supposed to be sufficiently smooth in (\ref{transf}), we cannot use the Heaviside step function $\Theta(x)$ to model the step distribution, since otherwise the result is incorrect. Also note that from (\ref{transf2}) 
\eq{\beta = \frac{1}{\epsilon} \lim_{x\to+\infty} \left(\int\limits_{-\infty}^{x}ds \; h(s) - x h(+\infty)\right)}

\end{itemize}

We can now calculate the evolution of physical quantities. We are mainly interested in the energy flow which is the off-diagonal component of the stress-energy tensor $T^{01}$, the correlation function of (primary) field operators and the entanglement entropy. The results obtained for the strip should be analytically continued to real time $\tau \to it $ and should give the correct asymptotic behaviour for times large in comparison with $m_{0}^{-1}(x)$ after the quench. 

In the next subsections we derive general formulae using the first approach. The derivation can be outlined as follows: we use the CFT transformation laws to express observables in the VWS in terms of those in the CWS in the form
\eq{\mathcal{O}_{vws}(\{w_{i}\}) = \mathcal{F}^{*}\Bigl[\{g(w_{i}),\overline{g(w_{i})}\},\mathcal{O}_{cws}(\{g(w_{i})\})\Bigr]}
or for infinitesimal transformations, if we keep only the first order in $f$ terms
\eqq[func]{\mathcal{O}_{vws}(\{w_{i}\}) = \mathcal{O}_{cws}(\{w_{i}\}) + \mathcal{F}\Bigl[\{f(w_{i}), \overline{f(w_{i})}\}\Bigr] \hat{\mathcal{D}} \mathcal{O}_{cws}(\{w_{i}\}) }
where the functionals $\mathcal{F}^{*}$ and $\mathcal{F}$ involve derivatives of the functions $g$ and $f$ respectively, as well as their complex conjugates, with $\mathcal{F}$ being linear in $f$. Also note that $w_{i}=x_{i}+i \tau_{i}$ and $\hat{\mathcal{D}}$ is a suitable differential operator. 

By the symmetry of the VWS under $z \to \bar{z}$ we deduce that 
\eqq[cond1]{\overline{g(w)} = g(\bar{w}) \qquad \mbox{ and similarly } \qquad  \overline{f(w)} = f(\bar{w})}
which is satisfied by the kernel (\ref{kernel}). This means that after the analytic continuation to real times $\tau \to it $, (\ref{func}) gives for the observables in the inhomogeneous problem
\eq{\mathcal{O}(\{x_{i},t_{i}\}) = \mathcal{O}_{0}(\{x_{i},t_{i}\}) + \mathcal{F}\Bigl[\{f(x_{i}-t_{i}),f(x_{i}+t_{i})\}\Bigr] \hat{\mathcal{D}} \mathcal{O}_{0}(\{x_{i},t_{i}\}) }
where the subscript 0 stands for the homogeneous expressions. This proves that it is only the restriction of $f$ along the real axis that we need in our calculation. As a last step we take the limit $\epsilon \to 0$, also using (\ref{transf2}) to express $f$ in terms of $h$.

\subsection{Energy flow}
%\emph{Derivation of the general formula for the energy flow in the case of small $h(x)$.}

Let us first calculate the energy flow. We will do this calculation in detail in order to demonstrate the procedure outlined above. Assuming that $f(z)$ is infinitesimal, the inverse map is to first order $w \to z = w-f(w)$ and the well known CFT formula for the transformation of the stress-energy tensor in complex coordinates gives
\begin{eqnarray}
\langle T_{vws}(w)\rangle & = (z'(w))^{2} \langle T_{{cws}}\rangle -\frac{c}{12}\{z,w\} \nn \\
& =  (1- 2 f'(w))\langle T_{cws}\rangle + \frac{c}{12} f'''(w)) 
\end{eqnarray}   
where 
\eq{\langle T_{cws} \rangle = \langle \overline{T}_{{cws}} \rangle =\frac{c}{24}\left(\frac{\pi}{2\epsilon}\right)^{2}} 
The energy flow is given by the expectation value of $ T^{01}(w,\bar{w})=i(T(w)-\overline{T(w)})/2\pi$ where we have taken into account the CFT normalisation factor $-1/2\pi$. Using the property (\ref{cond1}) that is satisfied by $f$ we find that
\eq{\langle\overline{T_{vws}(w)}\rangle = (1- 2 f'(\bar w))\langle \overline{T}_{{cws}}\rangle + \frac{c}{12} f'''(\bar w)) }
and combining the above results 
\eq{\langle T^{01}_{vws}(w,\bar{w}) \rangle = \frac{c i}{24 \pi} \left[\left(\frac{\pi}{2 \epsilon}\right)^{2}\Bigl(-f'(w)+f'(\bar{w})\Bigr)+f'''(w)-f'''(\bar{w})\right] } 
Now we substitute $w=x+i\tau$ and analytically continue to real times $\tau \to i t$ to obtain
\eqq[T]{\fl \langle T^{01}(x,t) \rangle = \frac{c}{24 \pi} \left[\left(\frac{\pi}{2\epsilon}\right)^{2}\Bigl(-f'(x-t)+f'(x+t)\Bigr)+f'''(x-t)-f'''(x+t)\right] }
Notice that, since $T^{\mu\nu}$ is not a scalar but a tensor, $T^{01}(x,t)$ in real time equals $ -i T^{01}(x,it)$ in imaginary time, that is apart from the substitution $\tau= it$ we also have to multiply by $dt/d\tau = -i$. Finally taking the limit $\epsilon \to 0$ and assuming that $h(x)$ varies slowly in comparison with $\epsilon$ so that we can use (\ref{transf2}), we end up with 
\eqq[T2]{ \langle T^{01}(x,t) \rangle = \frac{c \pi}{96 \epsilon^{3}} (h(x+t)-h(x-t))} 

From the last equation, it is apparent that the energy flows from the initial time hypersurface to both directions in a wave-like fashion with speed equal to 1 (the characteristic speed of the system, typically the speed of sound in condensed matter systems). The fact that it satisfies the wave equation is a general property of the stress-energy tensor in massless 2d theories, i.e. in CFT. What is non-trivial is the specific dependence on the initial distribution $h(x)$. 
% However since we know that the energy density satisfies the wave equation and that initially it is proportional to $h(x)$ and static (its time derivative is zero), we could directly derive the above equation for its evolution.

\subsection{Correlation functions}
% \emph{Derivation of the general formula for the correlation function.}

The next physically interesting quantity we can derive by the CFT transformation laws is the correlation function of a scalar primary field operator $\Phi$. If we define 
\eq{C(z_{1},z_{2})\equiv \langle \Phi(z_{1}) \Phi(z_{2})\rangle}
then its transformation law is
\eqq[Ftransf]{ C_{vws}(w_{1},w_{2}) = |w'(z(w_{1})) w'(z(w_{2}))|^{-\chi}C_{cws}(z(w_{1}),z(w_{2}))}
where $w_{j} = x_{j}+i \tau_{j}$ and $\chi$ is the scaling dimension of $\Phi$. Following the procedure described earlier we can write this transformation law in infinitesimal form. In particular to first order in $f$ we have $|w'(z(w))| = 1+ \mbox{Re} f'(w)$. The CWS correlation function $C_{cws}(z_{1},z_{2})$ has been found in \cite{cc-07}, but we can make a few more steps without using its explicit form, taking into account only the fact that it is invariant under space translations and interchange of the imaginary time variables. After some algebra we find that the equal time VWS correlation function to first order in $f$ is given by
\begin{eqnarray}
\fl C_{vws}(x_{1}+i\tau,x_{2}+i\tau) = \biggl( 1-\mbox{Re}\bigl[f(x_{1}+i\tau)-f(x_{2}+i\tau)\bigr]\frac{\partial}{\partial (x_{1}-x_{2})}- \nn \\
- \frac{1}{2}\mbox{Im}\bigl[f(x_{1}+i\tau)+f(x_{2}+i\tau)\bigr]\frac{\partial}{\partial \tau}- \biggr.   \nonumber \\ 
\biggl.-\chi\mbox{Re}\bigl[f'(x_{1}+i\tau)+f'(x_{2}+i\tau)\bigr]\biggr) C_{cws}(x_{1}+i\tau,x_{2}+i\tau) 
\end{eqnarray}
which after the analytic continuation to real times becomes
\begin{eqnarray}
\fl C(x_1,x_2,t)= & C_0(x_1-x_2,t)- \nonumber \\ 
&- \frac{1}{2}\frac{\partial C_0(x_1-x_2,t)}{\partial (x_1-x_2)} \left( f(x_1-t) + f(x_1+t) - f(x_2-t) - f(x_2+t) \right) + \nonumber \\
&+\frac{1}{4}\frac{\partial C_0(x_1-x_2,t)}{\partial t}  \left( f(x_1-t) - f(x_1+t) + f(x_2-t) - f(x_2+t) \right) - \nonumber \\
&-\frac{\chi}{2}C_0(x_1-x_2,t) \left( f'(x_1-t) + f'(x_1+t) + f'(x_2-t) + f'(x_2+t) \right)  \label{C00}
\end{eqnarray}
where the subscript 0 denotes the homogeneous case. Now we can use the explicit form of the homogeneous correlation function $C_0(r,t)$ 
\eqq[C0]{\fl
C_{0}(r,t) = \left[\left(\frac{\pi}{2\epsilon}\right)^{2} \frac{\cosh({\pi r}/{2 \epsilon})+\cosh({\pi t}/{\epsilon})}{8 \sinh^{2}({\pi r}/{4\epsilon})\cosh^{2}({\pi t}/{2\epsilon})}\right]^\chi \stackrel{\epsilon \to 0}{\sim} 
\cases{  
  e^{-{\chi \pi t}/{\epsilon}}   & \mbox{ if $t<r/2$,}  \\
  e^{-{\chi \pi r}/{2 \epsilon}}   & \mbox{ if $t>r/2$.} } }
take the limit $\epsilon \to 0$ and use (\ref{transf2}) to finally find
\begin{eqnarray}
\fl C(x_1,x_2,t)= C_0(x_1-x_2,t) \left[1 + \frac{\chi \pi}{4 \epsilon^2}  \left(\Theta(t-|x_1-x_2|/2) \int_{x_{2}-t}^{x_{1}-t} + \int_{x_{2}+t}^{x_{1}+t} ds \; h(s)  + \right. \right. \nn \\ 
 \left . + \Theta(|x_1-x_2|/2-t) \int_{x_{1}-t}^{x_{1}+t}+ \int_{x_{2}-t}^{x_{2}+t} ds \; h(s) \right)  - \nn \\
\left. -\frac{\chi}{2 \epsilon} \left( h(x_1-t) + h(x_1+t) + h(x_2-t) + h(x_2+t) \right) \right] \label{C}
\end{eqnarray}
where we assumed that $x_1>x_2$ without loss of generality. Although the physical significance of this relation will become transparent later when we apply it to concrete examples, it is already clear that the effect of the inhomogeneity of the initial state propagates in waves with unit speed and that this gives rise to different behaviour inside and outside the horizon at $t=|x_1-x_2|/2$.

\subsection{Entanglement entropy}
%\emph{Derivation of the general formula for the entanglement entropy.}

We now turn our attention to the evolution of the entanglement entropy whose calculation turns out to be an application of the results of the previous subsection, since it can be expressed in terms of a correlation function of primary operators too. The entanglement entropy between a subsystem $A$ defined by an interval $[x_{1},x_{2}]$ of length $l=|x_{1}-x_{2}|$ and the rest of the system, is equal to
\eqq[ent]{S_{A}= -\mbox{Tr}_{A} \rho_{A} \log \rho_{A} =-\left.\frac{\partial}{\partial n} \mbox{Tr} \rho_{A}^{n}\right|_{n=1}}
where $ \rho_{A} $ is the reduced density matrix of the subsystem $A$. 
From earlier work \cite{cc-04,cc-05} we know that $ \mbox{Tr} \rho_{A}^{n} $ turns out to be proportional to the correlation function of primary field operators $\Phi_{\pm n}$ defined on the strip geometry and having complex scaling dimensions $\Delta_{n}=\bar{\Delta}_{n}$
\eqq[trrho]{ \fl \mbox{Tr} \rho_{A}^{n} =c_{n} \langle \Phi_{n}(z_{1}) \Phi_{-n}(z_{2})\rangle = c_{n}
\left( \frac{|z_{1}-\bar{z}_{2}| |z_{2}-\bar{z}_{1}|} {|z_{1}-z_{2}| |\bar{z}_{1}-\bar{z}_{2}| |z_{1}-\bar{z}_{1}| |z_{2}-\bar{z}_{2}|} \right) ^{2n\Delta_{n}} } 
where $z_{1}, z_{2}$ correspond to the edges $x_{1},x_{2}$ of the interval and imaginary time $\tau$ and 
\eq{\Delta_{n} = \frac{c}{24} \left( 1-\frac{1}{n^{2}} \right) }
Notice that, since $\Delta_{1}=0$ and $\mbox{Tr} \rho_{A} = 1$, $c_{1}$ must be equal to 1. It should be mentioned that although (\ref{trrho}) is supposed to be valid for positive integer $n$ the analyticity of the expression allows us to calculate its derivative with respect to $n$ that appears in (\ref{ent}).  From (\ref{ent}), (\ref{trrho}) and the transformation law of correlation functions under conformal mappings $z \to w(z)$ which is 
\eqq[trPhi]{ \fl \langle \Phi_{n}(w_{1}) \Phi_{-n}(w_{2})\rangle = |w'(z(w_{1})) w'(z(w_{2}))|^{-2n\Delta_{n}}\langle \Phi_{n}(z(w_{1})) \Phi_{-n}(z(w_{2}))\rangle}
one can easily draw a general formula for the entanglement entropy in the inhomogeneous problem. Indeed, if we define the correlation function 
\eq{\fl \hat{C}(z_{1},z_{2}) \equiv \langle \Phi_{n}(z_{1}) \Phi_{-n}(z_{2})\rangle^{1/(2 n \Delta_{n})} =  \frac{|z_{1}-\bar{z}_{2}| |z_{2}-\bar{z}_{1}|} {|z_{1}-z_{2}| |\bar{z}_{1}-\bar{z}_{2}| |z_{1}-\bar{z}_{1}| |z_{2}-\bar{z}_{2}|}  }
then since $\hat{C}$ does not depend on $n$, the entanglement entropy is 
\eqq[ent3]{S_{A}=-\left.\frac{\partial}{\partial n} \left(c_{n}\hat C^{2n\Delta_{n}}\right)\right|_{n=1} = -\frac{c}{6}\log \hat{C} + \mbox{const.}}
On the other hand, from (\ref{trPhi}) we see that $\hat{C}$ itself transforms as 
\eqq[tr5]{ \hat C_{vws}(w_{1},w_{2}) = |w'(z(w_{1})) w'(z(w_{2}))|^{-1} \hat  C_{cws}(z(w_{1}),z(w_{2}))}
which is a special case of (\ref{Ftransf}) for $\chi = 1$. Hence in the infinitesimal inhomogeneous case $\hat{C}(x_1,x_2,t)$ is given by (\ref{C}) with $\chi = 1$. Substituting into (\ref{ent3}) and using the homogeneous form of the entropy already known from \cite{cc-05} 
\begin{eqnarray}
S_{A0}(l,t) = -\frac{c}{6}\log\left[\left(\frac{\pi}{2\epsilon}\right)^{2} \frac{\cosh({\pi l}/{2 \epsilon})+\cosh({\pi t}/{\epsilon})}{8 \sinh^{2}({\pi l}/{4\epsilon})\cosh^{2}({\pi t}/{2\epsilon})} \right] \sim \nn \\
\stackrel{\epsilon \to 0}{\sim} \frac{c}{3}\log{\epsilon} + 
 \cases{  
  \frac{c \pi t}{6 \epsilon}   & \mbox{ if $t<l/2$,}  \\
  \frac{c \pi l}{12 \epsilon}   & \mbox{ if $t>l/2$.} 
 } \label{S0}
\end{eqnarray}
we obtain the entanglement entropy after an inhomogeneous quench
\eem[S]{\fl S(x_{1},x_{2},t) = S_{0}(|x_{1}-x_{2}|,t) - \nn \\ 
- \frac{c\pi}{24\epsilon^{2}} \left[ \Theta(|x_{1}-x_{2}|/2-t) \left( \; \int\limits_{x_{1}-t}^{x_{1}+t} +  \int\limits_{x_{2}-t}^{x_{2}+t} ds \; h(s)\right) + \right. \nn \\
\left. + \Theta(t-|x_{1}-x_{2}|/2) \left(\;\int\limits_{x_{1}-t}^{x_{2}-t} +  \int\limits_{x_{1}+t}^{x_{2}+t} ds \; h(s) \right)\right] + \nn \\
+ \frac{c}{12 \epsilon}(h(x_{1}-t)+h(x_{1}+t)+h(x_{2}-t)+h(x_{2}+t))}

Similar comments like those for the correlation function apply here. The wave-like propagation of the entanglement and the different behaviour inside and outside the horizon will be fully explained later using the quasiparticle picture. 

\section{Application to the bump and step initial distributions \label{sec:3}}
%\emph{Calculation of the energy flow. Comments on the non-diffusive behaviour.
%Calculation of the correlation function and the entanglement entropy. Physical interpretation.
%Calculation of the entanglement entropy between the two semi-axes for arbitrarily large $h(x)$. Comparison with earlier work.}

Having found the general formulae we are now going to understand their characteristics. First of all, it is clear from (\ref{T2}), (\ref{C}) and (\ref{S}) that all quantities can be written as functions of the form $\; \mathrm{f}(x+t)\pm \mathrm{f}(x-t)$. This means that the inhomogeneity of the initial distribution evolves in the form of waves to both directions, verifying the physical picture of quasiparticles emerging from the initial time hypersurface. Any initial inhomogeneity located for example at the origin, will only affect a distant point $x$ after time $t=|x|$. This is another manifestation of the \emph{horizon} effect that we already mentioned and a direct consequence of causality. 

To make our results more transparent we apply them to two main cases, the bump and step distributions, emphasizing the latter. As models of those cases we should use smooth distributions localised for example at the origin and preferably such that $f(z)$ can be calculated in closed form at least from (\ref{transf2}). The parameter $\beta$ which corresponds to the size of the bump in the first case and $\alpha$ which is the height of the step in the second, must be both small if we wish $f(z)$ to be infinitesimal so that we can use (\ref{transf}). On the other hand, results that do not rely on this restriction can be obtained from the asymptotic approach using (\ref{asym}) and (\ref{gasym}). These will be valid away from the horizon lines $t=|x|$. The last approach is more useful for the qualitative descriptions of this section.

\

Let us start with the energy flow and assume for the moment that $f$ is infinitesimal. The meaning of (\ref{T2}) is almost obvious: the initial energy density distribution propagates as in the classical wave equation. Notice however that this relation is only valid for slowly varying distributions relative to $\epsilon$ and that in general the horizon will be smoothed over a distance of order $\epsilon$ in the CFT approach. 

To visualise the evolution we will describe what happens in the above two cases. In the case of the bump, the latter splits into two equal parts each of which moves to the two different directions. In the case of the step, the energy flow is non-zero only inside the horizon, that is for $t > |x|$, where it takes the constant value ${c\pi \alpha}/96 \epsilon^{3}$. Notice that if the strip width is larger on the right than on the left, which means that the opposite is true for the initial energy density, then after the quench the energy flows to the right as it should. Of course as we can check using the asymptotic approach, this behaviour is correct away from the horizon even for finite transformations.

\

Next comes the correlation function. We recall \cite{cc-06,cc-07} that in the homogeneous case (\ref{C0}) the correlation function $C_0$ decays exponentially in time until $t=r/2$ and then saturates to a value that depends exponentially on $r$ and we will use the infinitesimal corrections to $C$ (\ref{C}) along with the asymptotic form (\ref{asym}). 
Let us concentrate on the step distribution and consider two points $x_{1}$ and $x_{2}$ separated by some distance $r=|x_{1}-x_{2}|$, both lying on either the left or the right half of space, far away from the origin. 
If the pair of points is on the left then from (\ref{C}) and (\ref{asym}) we see that right after the quench, $C$ is equal to and evolves exactly like $C_{0}$ since all the corrections vanish. This is because the two points have not yet been affected by quasiparticles from the right half. This starts happening at $t=\min\{|x_{1}|,|x_{2}|\}$ and $C$ is changing until time equal to $r$ has passed. If now the pair of points is on the right half then right after the quench the correction terms in (\ref{C}) give exactly the first order corrections due to the substitution of $\epsilon$ in $C_{0}$ by $\epsilon + \alpha$. That is $C$ has the homogeneous form $C_{0}$ but with the local strip width $\epsilon + \alpha$ as expected since the two points are not affected by quasiparticles from the left half. As before this happens at $t=\min\{|x_{1}|,|x_{2}|\}$ and for time equal to $r$. After this time $C$ takes a value that is the same for the left and right half and equal to the average of the previous local values. To summarise, $C$ evolves initially like the homogeneous correlation function $C_{0}$ corresponding to the local strip width until it enters the horizon, when the contribution of quasiparticles from both halves mixes to the average of the left and right value.

\

A similar analysis holds for the entanglement entropy $S_A (t)$. In the homogeneous case (\ref{S0}) this increases linearly with time until $t=l/2$ when it saturates to a value proportional to the length $l$ of $A$. Notice that there is also some constant amount of entropy that depends only on $\epsilon$ in the form $\frac{1}{3} c \log{\epsilon}$. This offset is the part that is already there before the quench. Indeed since the theory is then massive the entropy according to \cite{cc-04} is $\frac{1}{3} c \log{\xi}$ where $\xi$ is the correlation length, $\xi \sim m_{0}^{-1} \sim \epsilon$. Moreover we mention that the entanglement in a massive theory is roughly speaking ``located'' close to the boundary points of $A$ over a distance $\xi$. 

\begin{figure}[htbp]
\begin{center}
\includegraphics[width=1.00\textwidth]{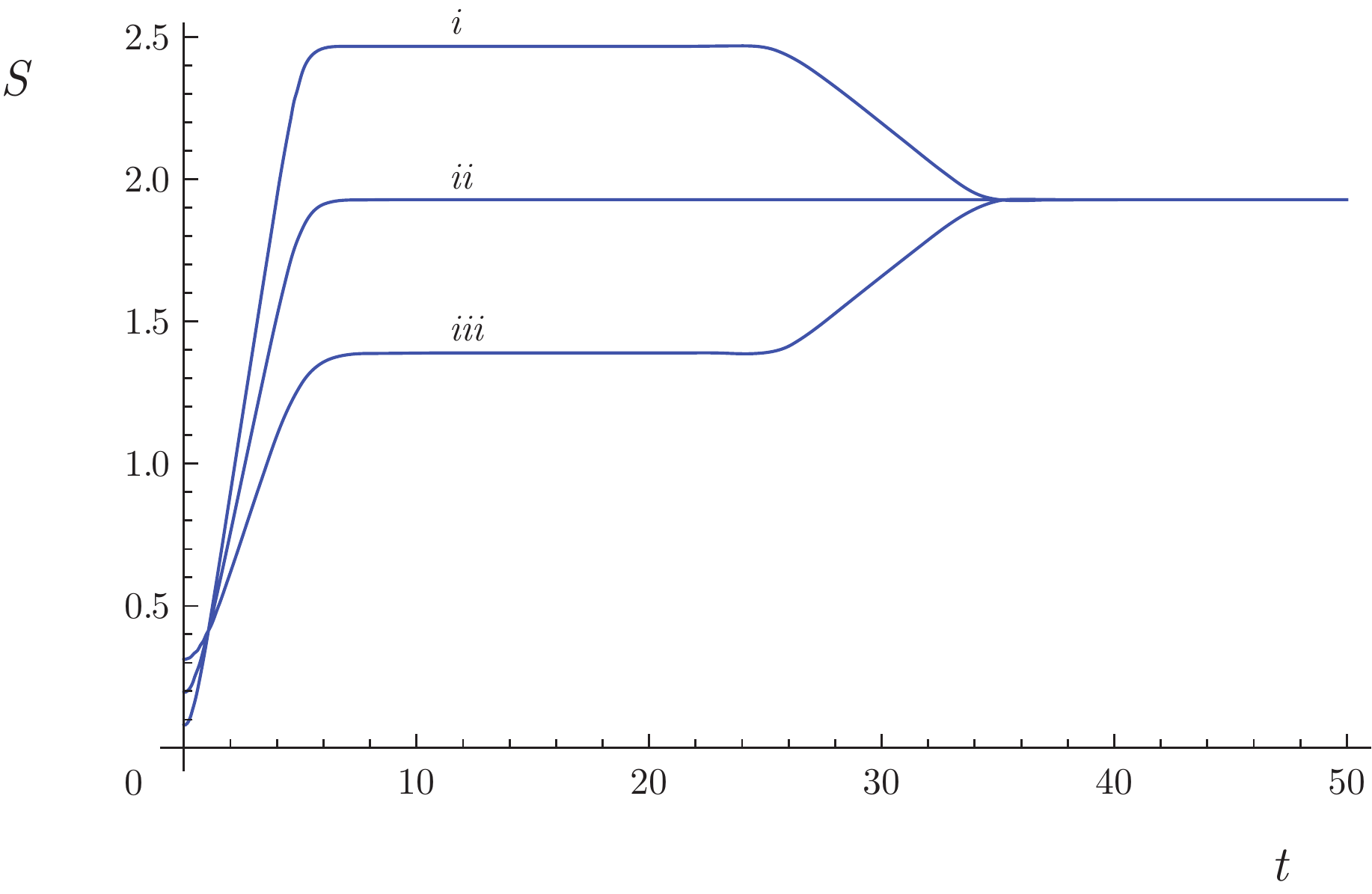}
\caption{Evolution of the entanglement entropy in the case of a step distribution, with the strip width smoothly varying over a distance of order $\epsilon=1$, from 2 on the left of the origin to 4 on the right. The actual transformation used is $f(z)=\frac{1}{2}\log(1+e^{2z})$ in which case the inverse transformation can be found analytically. The subsystem $A$ under consideration is of length $l=10$ and placed in 3 different positions with respect to the origin: with its middle at $x_{m}=-30$ (i), 0 (ii) and 30 (iii), all in units of $\epsilon=1$. We notice that in each case $S_{A}$ first saturates to the homogeneous saturation value that corresponds to the local (i,iii) or average width (ii), at time $t=l/2=5$. Note the different slopes and offset values before the saturation. In cases (i,iii) and at time $t=25$, i.e. the distance of the closest to the origin boundary of $A$, $S_{A}(t)$ starts changing again, since the first quasiparticles from the opposite half of space enter $A$. After time equal to $l=10$, the entropy saturates again to its final value which is common for all positions and equal to the homogeneous value that corresponds to the average width. In the plot we used (\ref{ent3}), (\ref{tr5}) and (\ref{F0d}) which are exact not only for infinitesimal transformations.}
\label{fig:ent}
\end{center}
\end{figure}

If we focus on the step distribution and consider again two cases for the position of the interval $A$, in the left or right half and away from the origin, then (\ref{S}) tells us that $S_{A}$ saturates first to the local homogeneous value at $t=l/2$, then it starts changing at time $t = \min\{|x_{1}|,|x_{2}|\}$ when the first quasiparticles from the opposite half enter $A$ and finally after time equal to $l$ it saturates again  to its final value
\eq{\frac{c}{3} \log \epsilon + \frac{\pi c l}{12 \epsilon}\left( 1-\frac{\alpha}{2 \epsilon} \right) + \frac{c}{6}\frac{\alpha}{\epsilon} }
This value is the same in both cases and exactly equal to the average of the initial saturation values on the left and on the right. It is also the same even if $A$ was in the middle containing the origin, as far as the length $l$ is the same (Fig. \ref{fig:ent}).

It is actually possible to go further and determine this asymptotic value of $S_{A}$ for arbitrary finite $h$ by using the general asymptotic form of the transformation $g$ (\ref{gasym}) together with (\ref{ent3}) and (\ref{tr5}). To this end we need however the expression for $\hat{C}_{cws}$ corresponding to different imaginary times because the points $x_{1}+i\tau$ and $x_{2}+i\tau$ will be mapped by $g$ to points with \emph{different} in general imaginary parts $\tau_{1}, \tau_{2}$ in the CWS. The required expression can be obtained from \cite{cc-07}
\ee{\fl \hat{C}_{cws}(r+i \tau_{1},i \tau_{2}) = \nn \\
\fl \left(\frac{\pi}{2\epsilon}\right)^{2} \frac{\cosh(\pi r/2 \epsilon)+\cos(\pi (\tau_{1}+\tau_{2})/2\epsilon)}{2 (\cos(\pi(\tau_{1}-\tau_{2})/2\epsilon)+\cos(\pi(\tau_{1}+\tau_{2})/2\epsilon)) (\cosh(\pi r/2\epsilon) - \cos(\pi(\tau_{1}-\tau_{2})/2\epsilon))}\label{F0d}} 
When we analytically continue to real times and take into account (\ref{gasym}) we find that in the limit  $t\to+\infty$
\numparts
\ee{
\fl r & \to \frac{1}{2} l \left(1+\frac{1}{1+{\alpha}/{\epsilon}}\right) \\
\fl \tau_{1}-\tau_{2}& \to \frac{1}{2 i} l \left(1-\frac{1}{1+{\alpha}/{\epsilon}}\right) \\
\fl \tau_{1}+\tau_{2}& \to \frac{1}{2 i} \left[ (x_{1}+x_{2})\left(1-\frac{1}{1+{\alpha}/{\epsilon}}\right) -2 t \left(1+\frac{1}{1+{\alpha}/{\epsilon}}\right) + \frac{2 \beta}{1+{\alpha}/{\epsilon}}\right]  }
\endnumparts
where $l=|x_{1}-x_{2}|$. Finally substituting into (\ref{F0d}), (\ref{ent3}) and (\ref{tr5}) we arrive at
\eq{\lim_{t\to\infty}S_{A}(l,t) = \frac{\pi c l}{12} \frac{1}{2}\left( \frac{1}{\epsilon}+\frac{1}{\epsilon+\alpha} \right) + \frac{c}{6}\left(\log \epsilon + \log(\epsilon + \alpha) \right)}
This leads to the important conclusion that the entropy finally saturates to a value independent of the position. It only depends on the length of the interval $l$, no matter whether it is in the left or right part or somewhere in the middle. As far as the initial conditions are concerned, it is completely determined by the two limits of $h(x)$ at $\pm \infty$, being the average of the corresponding homogeneous values. Bumps or any other characteristics of $h(x)$ do not affect the final saturation value.

\

Another interesting question is to find the entanglement entropy between the left and right halves in the case of a step initial distribution. Again we can do this for finite $\alpha$. Now that the subsystem $A$ is infinite, $\mbox{Tr}\rho^{n}_{A}$ transforms like $\langle \Phi_{n}(z) \rangle$ where $z$ corresponds to the edge of $A$ (i.e. the origin) and time $t$. This is because the other edge is at infinity and does not contribute to the correlation function $\langle \Phi_{n}\Phi_{-n}\rangle$. As before we can write $S_{A}=-(c/6)\log \hat{C}$ where $\hat{C}(z)\equiv\langle\Phi_{n}(z)\rangle^{1/(2n\Delta_n)}$ transforms as 
\eqq[12]{\hat{C}_{vws}(w)=|w'(z(w))|^{-1}\hat{C}_{cws}(z(w))}
and on the constant width strip takes the form found again in \cite{cc-07}
\eq{\hat{C}_{cws}(x+i\tau)=\frac{\pi}{4\epsilon}\frac{1}{\cos\frac{\pi\tau}{2\epsilon}}}
where as always we have modified the result for a strip centered in the middle. Setting $w=0+i\tau$ in (\ref{12}) and following the same procedure as before we finally find that for large times
\eqq[Sh]{S_{A}(t) \sim \frac{c \pi t}{12} \frac{1}{2} \left(\frac{1}{\epsilon}+\frac{1}{\epsilon + \alpha}\right) + \frac{c}{12}\left(\log \epsilon + \log(\epsilon + \alpha) \right)}
that is $S_{A}$ increases linearly in time with a rate that is the average of the homogeneous rates corresponding to the two different strip widths.

\section{Free field theory approach \label{sec:4}}
%\emph{Setup of the problem: massive (and massless) case, step initial distribution.}

Let us now study the inhomogeneous quench using a different approach: the real time evolution as prescribed by free field theory (FFT). This approach does not have the restriction that the hamiltonian after the quench be massless and it offers both a validity check of the CFT results in the massless case and an extension of the results to the massive case. As we will soon see, the time evolution of the two point correlation function can be easily found by solving the Heisenberg equations of motion. Then  the problem reduces to finding the initial correlation function, that is the correlation function in a theory with spatially inhomogeneous mass. This is formally equivalent to a scattering problem where the mass plays the role of the potential and it can be generally solved by means of perturbation theory, assuming that the spatial variation $\eta(x)$ of the mass is small relative to some characteristic value $m_{0}$. We are interested in the first order correction in $\eta$, which is sufficient in order to make a comparison with the conformal result. Unlike the CFT method where the correlation functions and the energy momentum tensor were derived independently, we will first calculate a general expression for the correlation function of the field and from this the energy flow. Then we explore the asymptotic behaviour of the latter in the special case of a step distribution in the deep quench limit. % In this special case an exact non-perturbative calculation is also possible if we use a sharp step distribution and find the Green's function for the field $\phi$ with appropriate boundary conditions at the origin. In the deep quench limit this solution leads to the same results.  

\subsection{Perturbative solution and calculation of the propagator}
%\emph{Derivation of the propagator and the energy flow. Comments on consistency with the CFT result. Comments on divergences.}

We start with the calculation of the two-point correlation function of the field operator $\phi$. The Heisenberg equation of motion for $\phi$ after the quench at $t=0$ is 
\eq{\ddot{\phi}(x,t) = ( \partial_{x}^2 -m^2 ) \phi(x,t)}
which can be readily solved in Fourier space where it reads
\eq{\ddot{\phi}(k;t) = -\omega_k ^2 \phi(k;t)}
with $\omega_k^2 = k^2 + m^2$ and $\phi(k;t)=\int{dx \, e^{-i k x} \phi(x,t)}$ the Fourier transform of $\phi(x,t)$. The solution is 
\eq{\phi(k;t) = \phi(k;0) \cos \omega_k t + \dot{\phi}(k;0) \frac{\sin \omega_k t}{\omega_k} }
so that the two-point correlation function is 
\begin{eqnarray}
\fl \langle \phi(x_1,t_1) \phi(x_2,t_2) \rangle =  \int{\frac{dk_1}{2\pi}\frac{dk_2}{2\pi} e^{ik_1 x_1+ik_2 x_2} 
\biggl(\langle \phi(k_1;0) \phi(k_2;0) \rangle \cos{\omega_1 t_1} \cos{\omega_2 t_2}  } + \nn \\
\fl + \langle \phi(k_1;0) \dot\phi(k_2;0) \rangle \cos\omega_1 t_1 \frac{\sin\omega_2 t_2}{\omega_2}  + \langle \dot\phi(k_1;0) \phi(k_2;0) \rangle \frac{\sin\omega_1 t_1}{\omega_1} \cos\omega_2 t_2 + \nn \\
+ \langle \dot\phi(k_1;0) \dot\phi(k_2;0) \rangle \frac{\sin\omega_1 t_1}{\omega_1} \frac{\sin\omega_2 t_2}{\omega_2}\biggr) \label{3p}
\end{eqnarray}
where $\omega_{i} \equiv \omega_{k_{i}}$. 

We have isolated the time evolution and we now need to calculate the correlation functions just before the quench, when the system lies on the initial ground state. The expectation values $\langle ... \rangle$ are therefore meant to be evaluated on this state so that we can forget the quench for the moment and focus on the field theory before that. If we write the initial mass distribution as $m_{0}(x)=m_{0}+\eta(x)$ and assume that $|\eta(x)| \ll m_{0}$ then the field equation before the quench is
\eq{(\partial_t^2-\partial^2_x+(m_{0}+\eta(x))^{2})\phi(x,t)=0}
or to first order in $\eta(x)$
\eqq[inhom-feq]{(\partial_t^2-\partial^2_x+m_{0}^{2}+2 m_{0}\eta(x))\phi(x,t)=0}
For later comparison with the CFT results, note that the correspondence between the strip width and the initial mass leads to $\epsilon = 1/m_{0}$ and so $h(x) = - \eta(x)/m_{0}^{2}$ where the variation is always assumed relatively small. 

The field equation above can be solved perturbatively in $\eta(x)$. The method is same as that used in the derivation of the Lippmann-Schwinger equation in scattering theory, with the inhomogeneous terms corresponding to the potential. The homogeneous field equation is the well-known Klein-Gordon equation
\eq{(\partial_t^2-\partial^2_x+m^2_{0} )\phi_{0}(x,t)=0}
whose retarded Green's function, satisfying
\eq{(\partial_t^2-\partial^2_x+m^2_{0} )G_{0R}(x,x',t,t')=-i\delta(x-x')\delta(t-t')}
is 
\eqq[G0R]{G_{0R}(x,x',t,t') = \Theta(t-t')(G_{0}(x-x',t-t')-G_{0}(x'-x,t'-t))}
with
\eqq[G0]{G_{0}(x-x',t-t')=\langle \phi_{0}(x,t) \phi_{0}(x',t') \rangle = \int{\frac{dk}{2\pi} e^{ik(x-x')-i\omega_{k}(t-t')} \frac{1}{2 \omega_{0k}}}}
where $\omega_{0k}=\sqrt{k^{2}+m_{0}^{2}}$. 
From these last relations, (\ref{inhom-feq}) can be written in the integral form 
\eq{\phi(x,t)=\phi_0(x,t)-2 i m_{0}\int {dt' \int{dx' G_{0R}(x,x',t,t') \eta(x')\phi(x',t')}}}
and since $\eta(x)$ is small this leads to first order to the solution 
\eq{\phi(x,t)=\phi_0(x,t)-2 i m_{0} \int {dt' \int{dx' G_{0R}(x,x',t,t') \eta(x')\phi_{0}(x',t')}} + \Or (\eta^{2})}
from which follows that before the quench
\begin{eqnarray} \fl
\langle \phi (x_{1},t_{1}) \phi (x_{2},t_{2}) \rangle = 
\langle \phi_{0} (x_{1},t_{1}) \phi_{0} (x_{2},t_{2}) \rangle - \nn \\ 
 -2 i m_{0} \int {dt' \int{dx' G_{0R}(x_{1},x',t_{1},t') \eta(x')\langle \phi_{0} (x',t') \phi_{0} (x_{2},t_{2}) \rangle}} - \nn \\ 
 - 2 i m_{0} \int {dt' \int{dx' G_{0R}(x_{2},x',t_{2},t') \eta(x')\langle \phi_{0} (x_{1},t_{1}) \phi_{0} (x',t') \rangle}} + \Or (\eta^{2})  
\end{eqnarray}
From (\ref{G0R}) and (\ref{G0}) we can work out the time integral (adding appropriate small imaginary numbers to the lower integration limits to ensure convergence). Passing to momentum space, the final result for the correlation function to first order in $\eta$ is 
\ee{\fl \langle\tilde\phi(k_1;t_1)\tilde\phi(k_2;t_2)\rangle = \nn \\
\fl = 2\pi \delta(k_{1}+k_{2}) \frac{e^{-i \omega_{01} (t_{1}-t_{2})}}{2 \omega_{01}}+\frac{2 m_{0} \tilde \eta(k_1+k_2) }{(\omega^2_{01}-\omega^2_{02})}\left(\frac{e^{-i \omega_{01} (t_{1}-t_{2})}}{2\omega_{01}}-\frac{e^{-i \omega_{02} (t_{1}-t_{2})}}{2\omega_{02}}\right) \label{Gk}}
where $\omega_{0i}\equiv \omega_{0k_{i}}$ and $\tilde \eta(k)$ is the Fourier transform of $\eta(x)$. 

Going back to (\ref{3p}), we see that we need the following equal time correlation functions between field and conjugate momentum operators $\pi = \dot{\phi}$, which can all be found from (\ref{Gk}) by applying appropriate time derivatives
\numparts
\ee{ \fl
\langle\tilde\phi(k_1;0)\tilde\phi(k_2;0)\rangle & = 2\pi \delta(k_{1}+k_{2}) \frac{1}{2 \omega_{01}}+\frac{2 m_{0} \tilde \eta(k_1+k_2) }{\omega^2_{01}-\omega^2_{02}}\left(\frac{1}{2\omega_{01}}-\frac{1}{2\omega_{02}}\right) \\ 
\fl \langle\tilde\phi(k_1;0)\dot{\tilde\phi}(k_2;0)\rangle & = \pi i \delta(k_{1}+k_{2}) \\ 
\fl \langle\dot{\tilde\phi}(k_1;0)\tilde\phi(k_2;0)\rangle & = - \pi i \delta(k_{1}+k_{2}) \\ 
\fl \langle\dot{\tilde\phi}(k_1;0)\dot{\tilde\phi}(k_2;0)\rangle & = 2\pi \delta(k_{1}+k_{2}) \frac{\omega_{01}}{2}+ \frac{2 m_{0} \tilde \eta(k_1+k_2)}{\omega^2_{01}-\omega^2_{02}}\left(\frac{\omega_{01}}{2}-\frac{\omega_{02}}{2}\right) \label{7.1}
}
\endnumparts
As a check, we can verify the consistency with the canonical commutation relations $ [ \tilde\phi(k_1;0), \dot{\tilde\phi}(k_2;0) ] = 2\pi i \delta(k_{1}+k_{2}) $. The next step is to substitute into (\ref{3p}) to find that the propagator after the quench is 
\begin{eqnarray} \fl
\langle \phi(x_1,t_1) \phi(x_2,t_2) \rangle = \nn \\ 
\fl = \int{\frac{dk}{2\pi} e^{ik (x_1-x_2)} 
\left( \frac{\omega_{0}}{2\omega^{2}} \sin\omega t_1 \sin\omega t_2 + \frac{1}{2\omega_{0}}\cos{\omega t_1} \cos{\omega t_2} - \frac{i}{2\omega}\sin\omega(t_{1}-t_{2}) \right) } +  \nn \\
\fl +  m_{0}\int{\frac{dk_1}{2\pi}\frac{dk_2}{2\pi} e^{ik_1 x_1+ik_2 x_2}\frac{\tilde\eta(k_1+k_2)}{\omega_{01}+\omega_{02}}\left(\frac{\sin\omega_1 t_1 \sin\omega_2 t_2}{\omega_1 \omega_2}  - \frac{\cos\omega_1 t_1 \cos\omega_2 t_2}{\omega_{01} \omega_{02}} \right)} \label{cor-f}
\end{eqnarray}
The first line is just the propagator in the homogeneous case found in \cite{cc-07} and the second line is the first order correction due to the inhomogeneity. In the deep quench limit $m_{0}\to \infty$, (\ref{cor-f}) can be written as 
\eqq[cor-f-dq]{\fl \langle \phi(x_1,t_1) \phi(x_2,t_2) \rangle =   \int{ \frac{dk_1}{2\pi} \frac{dk_2}{2\pi}\int{ds  e^{i k_1 (x_1-s)+i k_2 (x_2-s) } m_0(s)  \frac{\sin \omega_{1} t_1 \sin \omega_{2} t_2 }{2 \omega_{1} \omega_{2}}}} }
This expression can be found directly from (\ref{3p}) if we notice that the highest order in $m_{0}$ contribution comes from the initial momentum-momentum correlation function (\ref{7.1}) which can now be written in real space as $\langle  \dot{\phi}(x_1,0) \dot{\phi}(x_2,0)  \rangle \approx {m_0(x_{1})} \delta (x_1-x_2)/2 $. This is because the initial correlation functions fall exponentially fast over a distance of order $1/m_0 \to 0$ and so they can be approximated by $\delta$-functions with suitable coefficients which, in the current inhomogeneous problem, depend on the ``local mass'' $m_{0}(x)$. 

In the language of Feynman diagrams (\ref{cor-f-dq}) says that the dominant part of the correlation function between two points $(x_{1},t_{1})$ and $(x_{2},t_{2})$ comes from a diagram with two lines connecting the points with another one $(s,0)$ at the initial time hypersurface. What such a diagram represents physically is a pair of quasiparticles emerging from this hypersurface and reaching the two points. In general in order to cause correlations, two quasiparticles should emerge not necessarily from the same point, but from points separated by a distance of order $1/m_{0}(s)$, that is the ``local correlation length'', and the deep quench approximation consists in saying that these two points are so close that can be effectively identified.

In the massless case i.e. when $m=0$, $\omega_{k}=|k|$ and the Fourier transforms in (\ref{cor-f-dq}) can be calculated explicitly giving the correlation function as a convolution of the initial mass distribution
\eqq[cor-f-dq1]{\langle \phi(x_1,t_1) \phi(x_2,t_2) \rangle = \frac{1}{8} \int{ds \; m_0(s) \Theta(t_{1}-|x_{1}-s|) \Theta(t_{2}-|x_{2}-s|)} }
Notice that in the homogeneous case $m_{0} = \mbox{const.}$ and for equal times, we recover the deep quench massless propagator discussed in \cite{cc-07}
\eqq[cfhom]{\langle \phi(x_1,t) \phi(x_2,t) \rangle = 
 \cases{ 
0 & \mbox{ if $t<|x_{2}-x_{1}|/2$,} \\
(2t-|x_{2}-x_{1}|)m_{0}/8 & \mbox{ if $t>|x_{2}-x_{1}|/2$.} 
 } }

\begin{figure}[htbp]  % fix it
\begin{center}
\includegraphics[width=.80\textwidth]{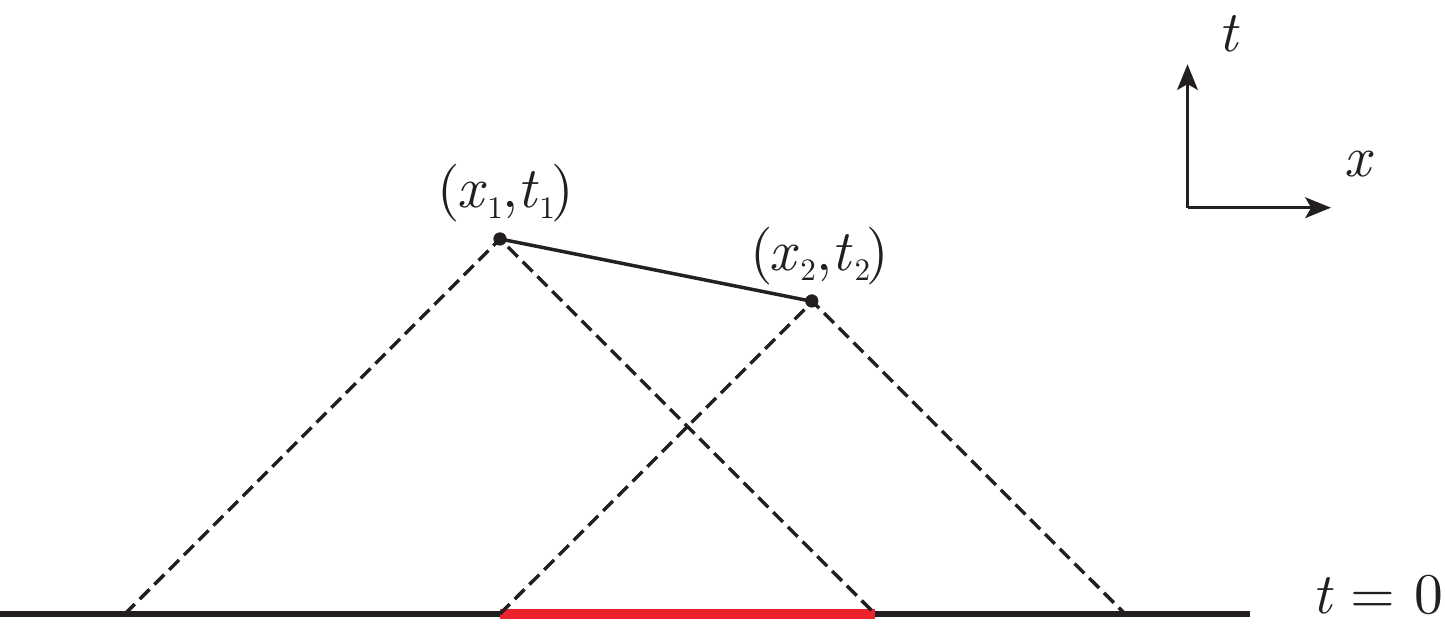}
\caption{Physical explanation of (\ref{cor-f-dq1}). The connected two-point correlation function is determined by the values of the initial distribution at the overlap of their horizons (red thick line).}
\label{fig:cf}
\end{center}
\end{figure}

The meaning of (\ref{cor-f-dq1}) is quite transparent: the correlation function between two spacetime points depends on the values of $m_{0}(s)$ only at those points of the initial time hypersurface that lie inside the horizons of both points (Fig. \ref{fig:cf}). 
The physical interpretation of correlations will be fully explained later using the quasiparticle picture. Also it will be shown that (\ref{cor-f-dq1}) is in agreement with the CFT formula (\ref{C}). 

\subsection{Energy flow}

The energy flow after the quench, i.e. the expectation value of the $T^{01}$ component of the stress-energy tensor, can be obtained by acting on the correlation function with $-\partial_{t_1}\partial_{x_2}$ and then setting $x_1=x_2$ and $t_1=t_2$
\eqq[Tff]{\langle T^{01} (x,t) \rangle = - \langle \pi(x,t) \partial_x \phi(x,t) \rangle = - \lim_{x'\to x} \lim_{t'\to t} \frac{\partial}{\partial t'} \frac{\partial}{\partial x} \langle \phi(x',t') \phi(x,t) \rangle}
From (\ref{cor-f}) we then find
\eqq[Tff2]{\fl T^{01}(x,t)= - m_{0} i \int{\frac{dk_1}{2\pi}\frac{dk_2}{2\pi} e^{i(k_1 +k_2) x} \frac{ \tilde \eta(k_1+k_2)}{\omega_{01}+\omega_{02}} k_{2} \omega_{1} \left(\frac{\cos\omega_1 t \sin\omega_2 t}{\omega_1 \omega_2} + \frac{\sin\omega_1 t \cos\omega_2 t}{\omega_{01} \omega_{02}} \right)}}
The homogeneous term, as expected, does not contribute to the energy flow. We are mainly interested in the value of the energy flow at the origin $x=0$ and for $t\to +\infty$. As before, in the massless case the expression above simplifies significantly
\eqq[Tff3]{\fl T^{01}(0,t) = - m_{0} i \int{\frac{dk_{1}}{2\pi}\frac{dk_{2}}{2\pi} \frac{\tilde \eta(k_1+k_2)}{2 (\omega_{01}+\omega_{02})}   \left(1 + \frac{k_{1} k_{2}}{\omega_{01} \omega_{02}} \right) \sin (k_1+k_2) t } }
but unlike before, we cannot use the deep quench limit in the first place as in (\ref{cor-f-dq}) as this would lead to ultraviolet divergences. In fact this should be expected since the double differentiation of the propagator dropped down a factor of $k_{2} \omega_{1}$ making $T^{01}$ more sensitive to ultraviolet divergences. Therefore the approximation that $m_{0} \gg k,q$ is not valid anymore. 

To extract the large $t$ behaviour of $T^{01}(0,t)$ we can probe the small frequency behaviour of its Fourier transform $\tilde T^{01}(0;\omega) = \int{dt \; e^{-i\omega t}} T^{01}(0,t) $ instead. Notice that the time integration runs from $-\infty$ to $+\infty$ although (\ref{Tff3}) is physically valid only for $t>0$. Since (\ref{Tff3}) is an odd function of $t$, $\tilde T^{01}(0;\omega)$ must be an odd function of $\omega$. This means that if $T^{01}(0,t)$ tends to a non zero value as $t\to +\infty$ then $\tilde T^{01}(0;\omega)$ must behave like $i/\omega$ for $\omega\to 0$, while if it tends to zero then $\tilde T^{01}(0;\omega)$ will have no singularity at $\omega= 0$. Setting $k_{1}=k+q$ and $k_{2}=k-q$ 
\begin{eqnarray}
\fl \tilde T^{01}(0;\omega)  = - m_{0} i \int{\frac{dk}{2\pi}\frac{dq}{2\pi} \frac{\tilde \eta(2k)}{(\omega_{0(k+q)}+\omega_{0(k-q)})}  \left( 1 + \frac{k^{2}-q^{2}}{\omega_{0(k+q)}\omega_{0(k-q)}} \right)  \int{dt e^{-i\omega t}}  \sin 2k t } \nn \\
\fl =  - m_{0} \int{\frac{dk}{2\pi}\frac{dq}{2\pi} \frac{ \tilde \eta(2k)}{2 (\omega_{0(k+q)}+\omega_{0(k-q)})}  \left( 1+\frac{k^{2}-q^{2}}{\omega_{0(k+q)}\omega_{0(k-q)}} \right) 2 \pi(\delta(2k-\omega)-\delta(2k+\omega))} \nn \\
\fl =  - \frac{1}{4} m_{0} (\tilde \eta(\omega)-\tilde \eta(-\omega)) \int{\frac{dq}{2\pi} \frac{1}{(\omega_{0(\omega/2+q)}+\omega_{0(\omega/2-q)})}  \left(1 + \frac{\omega^{2}/4-q^{2}}{\omega_{0(\omega/2+q)}\omega_{0(\omega/2-q)}} \right) }
\end{eqnarray}
and for $\omega \to 0$ 
\eq{\fl \tilde T^{01}(0;\omega) \sim - \frac{1}{8} m_{0} (\tilde \eta(\omega)-\tilde \eta(-\omega)) \int{\frac{dq}{2 \pi} \frac{m^{2}_{0}}{\omega^{3}_{0q}}  = - \frac{1}{8\pi} m_{0} (\tilde \eta(\omega)-\tilde \eta(-\omega)) }}
Applying our result to the case when the initial mass distribution is a step function of step $a$, where $\tilde \eta (k) = a i/k$ we find
\eqq[Tomega]{ \tilde T^{01}(0;\omega\to 0) \sim - \frac{m_{0} a i }{4 \pi \omega}   }
which, according to our discussion above, means that the energy flow at the origin tends for large times to a non-zero value, more specifically to $-a m_{0}/(8\pi)$. This result proves once again the non-diffusive behaviour of the energy flow. Notice that if $a >0$, i.e. the initial mass and consequently the initial energy density is higher in the right half of space than it is in the left, then the minus sign indicates that the energy flows from right to left, as it is supposed to do. 

The massive case is also interesting independently from the massless one since thermalisation occurs for a different reason in the two cases. Following a similar calculation presented in \ref{App3}
we conclude again that the energy flow at $t\to +\infty$ tends to a non-zero value.

\section{Comparison between the conformal and free field theory results \label{sec:5}}

Since we were able to calculate the energy flow and  correlation function both in the CFT and FFT approach, we can now compare our results. We start with the energy flow. In both cases we have shown that it does not decrease with time inside the horizon. If we compare the asymptotic values at the origin for large times using the correspondence relations $\epsilon = 1/m_{0}$ and $h(x)=-\eta(x)/m_{0}^{2}$ we notice that the two results agree up to a numerical factor of $\pi^{2}/12$. Consequently the CFT method is consistent with FFT as far as the qualitative behaviour of the energy flow is concerned, the only difference being in numerical factors. This remark holds in the homogeneous case as well if one calculates the energy density instead, as explained in \ref{App4}. 

Now for the correlation function we have to take into account that in the gaussian model (free boson) the primary field $\Phi$, whose correlation function $C(x_{1},x_{2},t)$ is given by the CFT expression (\ref{C}), is not the field $\phi$ of FFT but its imaginary exponential. More specifically $C(x_{1},x_{2},t)$ should be compared to $\langle e^{iq\phi(x_{1},t)} e^{-iq\phi(x_{2},t)} \rangle$ where $q$ is an arbitrary constant. Using properties of gaussian integrals we have
\ee{ \langle e^{iq\phi(x_{1},t)} e^{-iq\phi(x_{2},t)}  \rangle = e^{- \frac{q^{2}}{2}\langle(\phi(x_{1},t)-\phi(x_{2},t))^{2}\rangle} = \nn \\
= \exp{\left[-\frac{q^{2}}{2}(G(x_{1},x_{1},t)+G(x_{2},x_{2},t)-2G(x_{1},x_{2},t))\right]} }
where $G(x_{1},x_{2},t)$ is the correlation function of $\phi$ in the massless case, as given by (\ref{cor-f-dq1}). After some algebra we find that to first order in $\eta(x)$
\begin{eqnarray}
\fl \langle e^{iq\phi(x_{1},t)} e^{-iq\phi(x_{2},t)}  \rangle = e^{{q^{2}}(G_0(x_{1}-x_{2},t)-G_0(0,t))} \Biggl[1 - \frac{q^2}{16} \int\limits_{-\infty}^{+\infty}{ds \; \eta(s) }\times \nn \\
\fl \times {\Bigl( \Theta(t-|x_{1}-s|)+\Theta(t-|x_{2}-s|) -2\Theta(t-|x_{1}-s|) \Theta(t-|x_{2}-s|)\Bigr)}\Biggr] \label{FFTcf1}
\end{eqnarray}
where $G_0(x_{1}-x_{2},t)$ is the homogeneous correlation function (\ref{cfhom}). 

On the other hand, the CFT result (\ref{C}) can also be written in a comparable form 
\eqq[CFTcf2]{
C(x_1,x_2,t)= C_0(x_1-x_2,t) \left[1+ \frac{\chi \pi}{4 \epsilon^2} \int\limits_{-\infty}^{+\infty} ds \; h(s) K(x_1,x_2,t;s) \right] }
where the kernel $K(x_1,x_2,t;s)$ is
\ee{ \fl K(x_1,x_2,t;s) = \Theta(|x_1-x_2|/2-t) \Bigl(\Theta(t-|x_{1}-s|)+\Theta(t-|x_{2}-s|)\Bigr) + \nn \\
\fl + \Theta(t-|x_1-x_2|/2)\Bigl(\Theta(s-x_2+t)\Theta(-s+x_1-t)+\Theta(s-x_2-t)\Theta(-s+x_1+t)\Bigr) }
and we kept only first order in $\epsilon$ terms. The homogeneous correlation functions $\exp{[{q^{2}}(G_0(x_{1}-x_{2},t)-G_0(0,t))]}$ and $C_0(x_1-x_2,t)$ have been already shown to be equal \cite{cc-07} assuming that $q^2=4\chi\pi$. What we wish to verify now is that the first order corrections agree too. This is true since the integration kernels in (\ref{FFTcf1}) and (\ref{CFTcf2}) are in fact identical and even the numerical coefficients are equal if we use the previous substitution for $q$.

\section{Discussion and conclusions. \label{sec:6}}
%\emph{In this section we should emphasize the physical interpretation using the concept of semiclassical quasiparticles and compare our results with the thermal case.}

\subsection{The quasiparticle interpretation}

Our results can be easily interpreted using the physical picture of quasiparticles developed in earlier work \cite{cc-07} and verified for the special case of a domain wall initial distribution in \cite{chd-08}. This can be seen in the formula (\ref{S}) that gives the evolution of the entanglement entropy for example. 

\begin{figure}[htbp]
\begin{center}
\includegraphics[width=.80\textwidth]{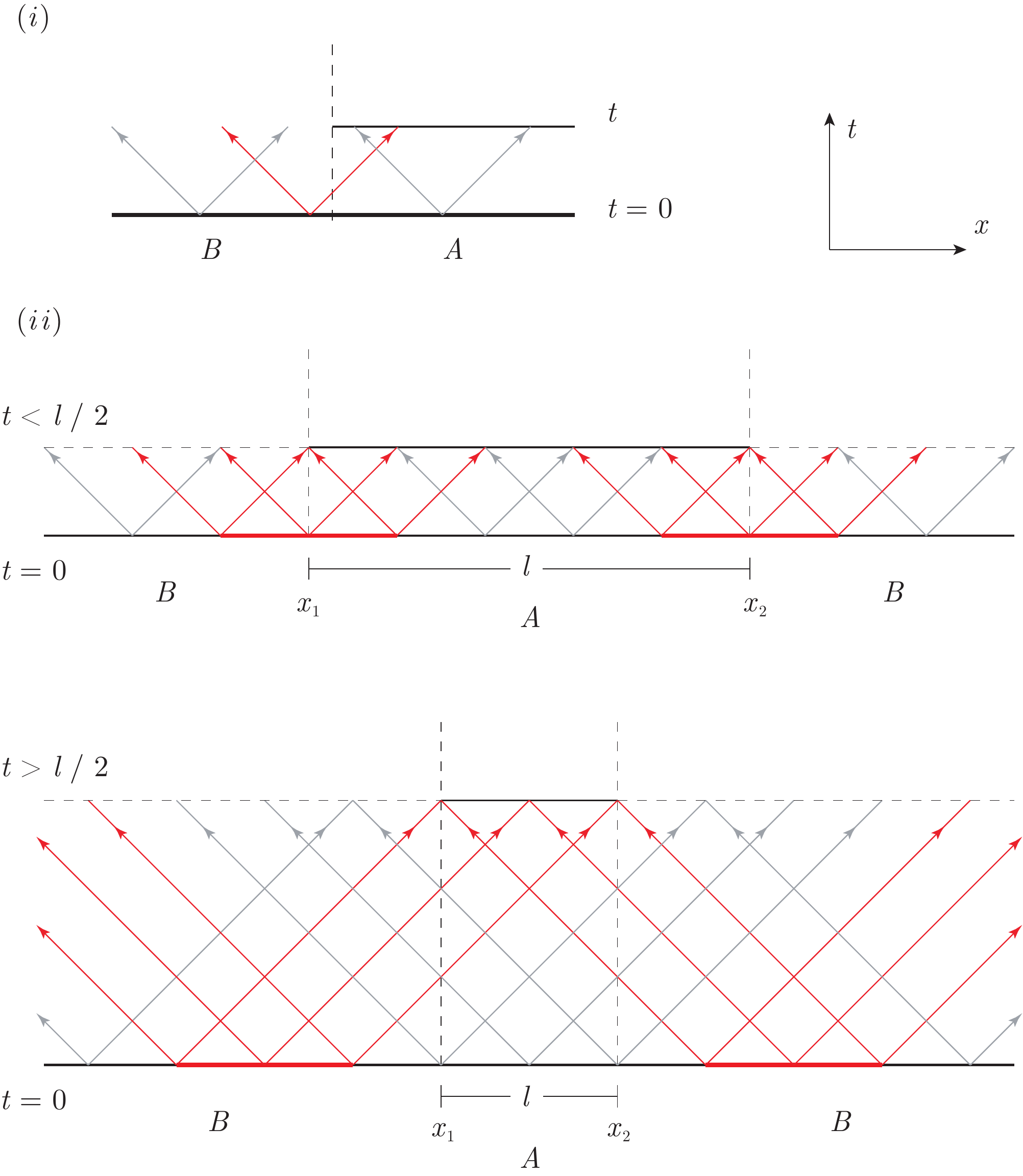}
\caption{Illustration of the physical interpretation of the entanglement entropy evolution using the concept of entangled quasiparticles. 
\begin{enumerate}
\item ~Pairs of entangled quasiparticles emitted from the same point on the $t=0$ hypersurface. One of the quasiparticles of the pair denoted by dark red colour is inside subsystem $A$ at some time $t$, while the other is in its complement $B$. Therefore this pair contributes to the entanglement  $S_{A}(t)$ between $A$ and $B$ at that time. In contrast, pairs like those denoted by light grey colour whose quasiparticles are both in the same subsystem, either $A$ or $B$, do not contribute to $S_{A}(t)$. 
\item ~The red thick lines denote the regions on the $t=0$ hypersurface where entangled quasiparticles that contribute to $S_{A}(t)$ come from. For $t<l/2$ these regions are $[x_{1}-t,x_{1}+t]$ and $[x_{2}-t,x_{2}+t]$, while for $t>l/2$ they are $[x_{1}-t,x_{2}-t]$ and $[x_{1}+t,x_{2}+t]$.
\end{enumerate}
}
\label{fig:qp}
\end{center}
\end{figure}

First we repeat how this physical picture applies to the homogeneous case. As always we consider a subsystem $A$ which is an interval of length $l$ while the complement is the subsystem $B$. Then the entanglement entropy between the two subsystems after the quench is given by (\ref{S0}) up to an additive constant independent of $\epsilon$. The constant value $\frac{1}{3}c\log{\epsilon}$ is the part that corresponds to the massive field theory before the quench. In that case correlations are restricted to between points separated by distances of order $1/m_{0} \sim \epsilon $ which is the correlation length. Hence the entanglement between $A$ and $B$ is only due to correlations between points close to the two boundaries of $A$. Since the initial state has energy density much higher than the ground state of the Hamiltonian after the quench, it acts as a source of quasiparticles moving with unit speed to both directions. Two quasiparticles emitted from the same initial point are entangled and if they move to different directions they can reach the two different subsystems increasing the entanglement between them. The latter must be at any time proportional to the number of such entangled pairs which is simply proportional to the length of the region $\mathcal{E}(A)$ where these pairs are emitted from (Fig. \ref{fig:qp}). 
This is the total length of the left and right projections of $A$ on the initial line apart from their intersection, since quasiparticles coming from there both end up inside $A$. Also the number of quasiparticle pairs emitted from some point must be proportional to the initial energy density at this point, which in the homogeneous case is constant and proportional to $1/\epsilon$. 

Keeping all these in mind, we can easily see that shortly after the quench, the pairs that contribute to the entanglement are emitted from within distance equal to $t$ from the boundaries of $A$. This causes a linear increase $\sim 4t/\epsilon$. However at time $t=l/2$, the region $\mathcal{E}(A)$ reaches its maximum length $2l$, since the left and right projections of $A$ on the initial line no longer overlap. This explains the saturation to a value $\sim 2l/\epsilon$. By comparison with (\ref{S0}) the proportionality factor turns out to be equal to $c\pi/24$. 

Following the same arguments it should be straightforward to generalise to the inhomogeneous problem. From (\ref{S}) the part of the entanglement entropy that corresponds to the massive theory before the quench is 
\eq{S(x_{1},x_{2},0)=\frac{c}{3}\log{\epsilon}+\frac{c}{6\epsilon}(h(x_{1}) + h(x_{2}))}
This is simply the sum of the contributions of the two boundaries, if we take into account that now it is the \emph{local} correlation length at each point that should be used 
\eq{\frac{c}{6} [\log(\epsilon+h(x_{1}))+\log(\epsilon+h(x_{2}))] }
and that $h(x) \ll \epsilon$. To calculate the contribution of this initial entanglement after the quench, we must take into account that due to the wave propagation, $A$ is affected not by the initial value at $x_{1}, x_{2}$ but at their projections $x_{1}\pm t, x_{2}\pm t$. This justifies the term
\eq{\frac{c}{12 \epsilon}(h(x_{1}-t)+h(x_{1}+t)+h(x_{2}-t)+h(x_{2}+t))}
Now we calculate the contribution of the quasiparticles. As before, until $t=l/2$, this comes from the intervals $[x_{1}-t,x_{1}+t]$ and $[x_{2}-t,x_{2}+t]$, but now the emission rate at some point $s$ is proportional to the local initial energy density $\sim 1/(\epsilon+h(s))$. This leads to
\eq{\frac{c\pi}{24} \left( \; \int\limits_{x_{1}-t}^{x_{1}+t} ds \frac{1}{\epsilon + h(s)} +  \int\limits_{x_{2}-t}^{x_{2}+t} ds \frac{1}{\epsilon + h(s)}\right)}
and since $h(x) \ll \epsilon$
\eq{\frac{c\pi t}{6\epsilon} - \frac{c\pi}{24\epsilon^{2}} \left( \; \int\limits_{x_{1}-t}^{x_{1}+t} ds \; h(s)+  \int\limits_{x_{2}-t}^{x_{2}+t} ds \; h(s) \right)}
The first term can be recognised as the homogeneous part while the rest is due to the inhomogeneity. 
Similarly after $t=l/2$ we find 
\eq{\frac{c\pi l}{12\epsilon} - \frac{c\pi}{24\epsilon^{2}} \left( \; \int\limits_{x_{1}-t}^{x_{2}-t} ds \; h(s)+  \int\limits_{x_{1}+t}^{x_{2}+t} ds \; h(s) \right)}
Adding altogether we obtain (\ref{S}). 

A similar analysis applies to the expression for the correlation function $C(x_{1},x_{2},t)$. This shows that the quasiparticle interpretation successfully explains all the details of the evolution of correlations following an inhomogeneous quantum quench. This picture should also be valid in the massive case, the difference being that the quasiparticles now have finite lifetimes and propagate with various velocities up to the maximum one \cite{cc-07}. This smoothes out the horizon and leads to spatial oscillations of correlations inside and exponential decay outside it, as can be shown by stationary phase arguments.

\subsection{Comparison to the thermal analogue}

Our findings enable us to answer the main question that motivated the study of this problem, that is if there are any similarities between an inhomogeneous quantum quench and its thermal analogue. This is the evolution of an inhomogeneous initial temperature distribution $u_{0}(x)$ which is given by the heat equation $\partial_{t}u(x,t) = \kappa \partial^{2}_{x}u(x,t)$ with initial condition $u(x,0)=u_{0}(x)$. The solution is 
\eq{u(x,t)= \frac{1}{\sqrt{4 \pi \kappa t}} \int\limits_{-\infty}^{+\infty}{ds \, u_{0}(s)\exp{[-{(x-s)^2}/{4 \kappa t}]}}}
from which we can find the heat flow by Fourier's law
\eq{\fl j_{q}(x,t)\equiv \frac{dQ}{dt} = - \lambda \partial_{x} u(x,t) = - \frac{\lambda}{\sqrt{4 \pi \kappa t}} \int\limits_{-\infty}^{+\infty}{ds \, u'_{0}(s)\exp{[-{(x-s)^2}/{4 \kappa t}]}}}
As can be seen from the last equation and mentioned in the introduction, for any step-like initial distribution the heat flow at the origin decreases as $1/\sqrt{t}$ for large times
\eq{j_{q}(0,t ) \sim - \frac{\lambda \alpha}{\sqrt{4 \pi \kappa t}} }
where $\alpha$ is the size of the step. This can be considered as a characteristic of diffusive behaviour thus providing a simple test for our analysis. 

In a quantum quench on the other hand, the energy density which is supposed to be proportional to the effective temperature, if we could assign a local meaning to it, exhibits wave-like non-diffusive behaviour instead as we can see from (\ref{T2}). The same message comes from (\ref{Tomega}) and (\ref{Tmass}) when we see them through the lens of the aforementioned test. We thus conclude that the effective temperature is not meaningful as a local quantity. Notice that the non-decreasing of the energy flow is also true in the massive case $m \neq 0$, even though the quasiparticles have finite lifetimes. Recall that for $m \neq 0$ it is the bosonic propagator itself that thermalises, unlike the conformal case where thermalisation occurs on the level of correlation functions of primary field operators. Thus the nature of thermalisation is qualitatively different in each of these two cases and one should consider it as an independent effect. 

Another aspect of the comparison is whether the entanglement entropy $S_{ent}$ resembles the thermodynamic one $S_{th}$. We already know from the homogeneous case that $S_{ent}$ becomes extensive when it saturates and of course the same holds in the present inhomogeneous case. One might be tempted then to define an entanglement entropy current and ask if this plays a role similar to the heat current $j_{q} = T (dS_{th}/dt)$ where $T$ is the temperature. Once again this question is simplified for a step distribution and with the complementary subsystems $A$ and $B$ between which $S_{ent}$ is measured, being the two halves of space, on the left and right of the origin. From (\ref{Sh}) we see that in CFT the entanglement entropy rate for large times is constant, not decreasing as it should happen if it exhibited diffusive behaviour like the heat current.  % open problem: j_{q}=j_{e}-\mu j_{n}

As a final remark we will discuss the second interesting question arisen in the introduction which was what information about the initial state survives in the stationary values of local observables for large times. As must be clear from the CFT expressions (\ref{T2}), (\ref{C}) and (\ref{S}) in the limit $t \to +\infty$ the only relevant parameters of the initial distribution $h(x)$ are its asymptotic values in the limits $x\to \pm \infty$. This must be obvious considering the wave-like nature of the evolution. We should mention by the way that, even if the evolution was of diffusive nature, those two limits would still be the only relevant parameters determining the large time behaviour. For example the uniform relaxation temperature is simply the average of the initial temperature distribution $\int{ds \; u_{0}(s)}$ which equals the average of the above two limits $u_{0}(\pm \infty)$.

\ack

This work was supported in part by EPSRC grant EP/D050952/1. S. Sotiriadis acknowledges financial support from St John's College, Oxford, and the A.G.Leventis Foundation. He would also like to thank Fabian Essler for useful discussions and Pasquale Calabrese for reading the final version of the paper.

\appendix

\section{Derivation of the conformal map \label{App1}}

We seek a conformal transformation $g(w)$ from the strip with variable width $2(\epsilon + h(x))$ to the strip with constant width $2\epsilon$, which means that $g$ must satisfy the boundary conditions (\ref{bc1}). 
As mentioned in the main part of the article, these conditions have the disadvantage of being defined along the boundary of the VWS whose shape is nontrivial. Let us define the inverse transformation $ 1 + f \equiv g^{-1}  $ from the CWS to the VWS. If we expressed the problem in terms of $f$ then the boundary conditions would be defined along straight lines, but (\ref{bc1}) cannot be translated into sufficient boundary conditions for $f$ unless we assume that the transformation is infinitesimal. More specifically, if $h(x)$ is small compared to $\epsilon$ and sufficiently smooth, then $g$ is close to the identity and $f$ is of the order of $h$ so that the boundary condition implied for $f$ to first order in $h$ is (\ref{bcf}) $\mbox{Im}f(x\pm i \epsilon) = \pm h(x)$.

Now the problem reduces to finding the analytic function $f(z)$ defined on the CWS and obeying (\ref{bcf}). $\mbox{Im}f$ is a solution of the Laplace equation which ensures that the boundary conditions above along with conditions at $\pm \infty$ are sufficient for determining $\mbox{Im}f$, while $\mbox{Re}f$ can be determined using the Cauchy-Riemann equations up to a real additive constant. This is irrelevant for our purposes since we can check that it would vanish from all of the expressions for the physical observables that we are interested in. Hence we can fix its value by requiring that $f$ is equal to zero for vanishing $h$. Without loss of generality we suppose that $h(-\infty)=0$ and then
\eqq[lim]{\lim_{{Re}z \to -\infty} f(z) = 0} 

\begin{figure}[htbp]
\begin{center}
\includegraphics[width=.80\textwidth]{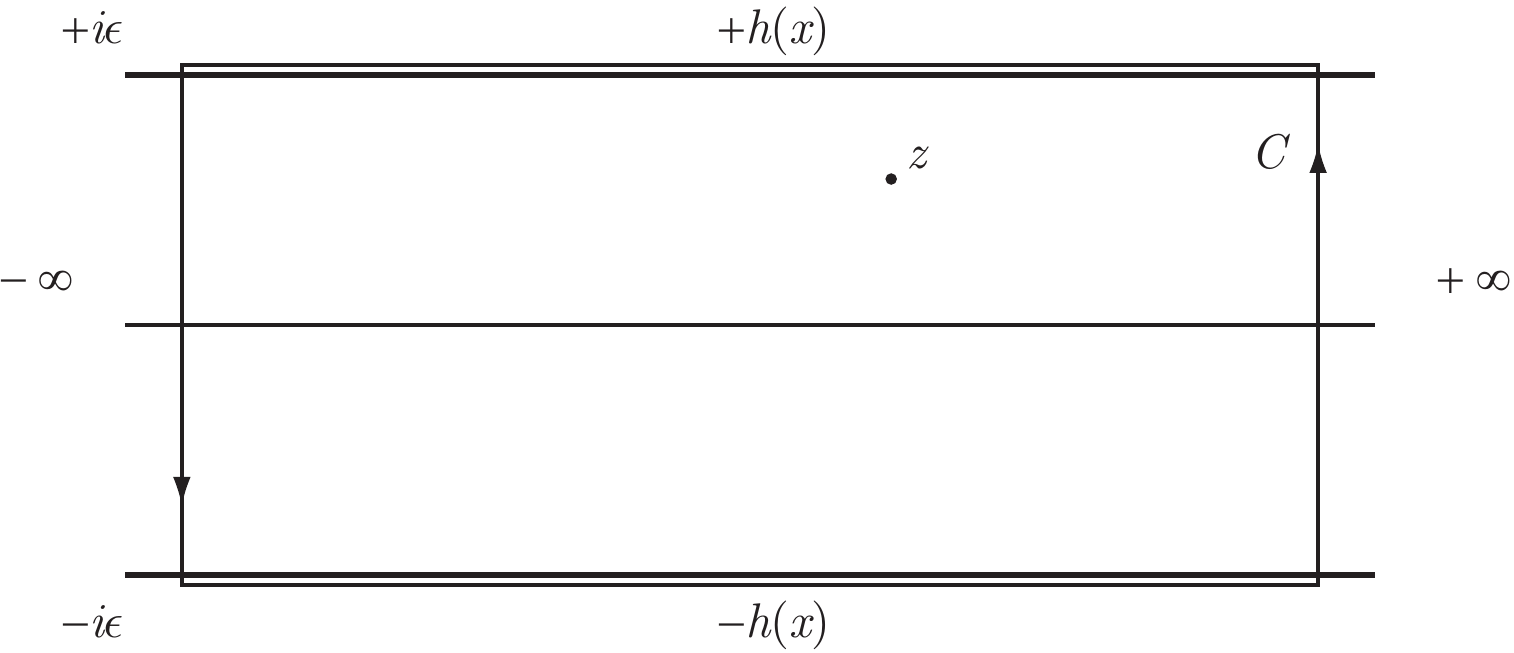}
\caption{Contour integration for the derivation of  $f(z)$.}
\label{fig:contour}
\end{center}
\end{figure}

One way to find $f(z)$ is using a generalised form of Cauchy's integral formula with an appropriate kernel. In particular we have
\eq{f(z)=\frac{1}{2\pi i} \oint_{C} D(z'-z)f(z')dz'}
where $C$ is a closed contour around $z$ and $D(z'-z)$ is any function with a simple pole with unit residue at $z'=z$ as the only singularity within $C$. If we choose $C$ to go along the boundaries of the strip (Fig.\ref{fig:contour}) then 
\ee{\fl f(z)=\frac{1}{2\pi i} \int\limits_{-\infty}^{+\infty} \left(D(s-i\epsilon-i\eta-z)f(s-i\epsilon)-D(s+i\epsilon+i\eta-z)f(s+i\epsilon)\right)ds + \nn \\ 
\fl + \frac{1}{2\pi } \int\limits_{-\epsilon}^{+\epsilon} \left(D(+\infty + i s -z)f(+\infty + i s )-D(-\infty + i s -z)f(-\infty + i s )\right)ds }
where $\eta \to 0+$ simply serves as a reminder of the prescription that the pole is always enclosed inside the contour, even for $z$ on the boundary. Using (\ref{lim}) and the symmetry of the problem under reflections $z\to \bar z$ which implies that $f(\bar z)=\overline{f(z)}$, we find
\ee{\fl f(z)=\frac{1}{2\pi i} \int\limits_{-\infty}^{+\infty} \left(D(s-i\epsilon-i\eta-z)-D(s+i\epsilon+i\eta-z)\right)\mbox{Re}f(s+i\epsilon)ds \, - \nn \\ 
\fl - \frac{1}{2\pi} \int\limits_{-\infty}^{+\infty} \left(D(s-i\epsilon-i\eta-z)+D(s+i\epsilon+i\eta-z)\right)\mbox{Im}f(s+i\epsilon)ds \, + \nn \\ 
\fl + \frac{1}{2\pi } \int\limits_{-\epsilon}^{+\epsilon} D(+\infty + i s -z)f(+\infty + i s )ds }
The last relation suggests that we choose $D(z)$ such that $D(z+2i\epsilon)=D(z)$ and $\lim_{{Re}z \to +\infty} D(z) = 0$, because then $f(z)$ is completely determined by the boundary condition (\ref{bcf}) 
\eq{f(z)= - \frac{1}{\pi} \int\limits_{-\infty}^{+\infty} D(s+i\epsilon-z) h(s) ds }
Now it is relatively easy to see that one function that satisfies all the above conditions is 
\eq{D(z) = \frac{\pi}{\epsilon} \frac{1}{(e^{\pi z/\epsilon}-1)} }
which finally gives 
\eq{f(z)= \frac{1}{\epsilon} \int\limits_{-\infty}^{+\infty} \frac{1}{e^{-\pi (z-s)/\epsilon}+1} h(s) ds }
i.e. (\ref{transf}) and (\ref{kernel}).

\section{Free field theory calculation of the energy flow in the massive case}\label{App3}

\

We will calculate the large time asymptotic behaviour of $T^{01}(x,t)$ at $x=0$ and with $m \neq 0$ following the same method as for the massless case, that is using its Fourier transform with respect to time $\tilde T^{01}(0;\omega)$. As before, if this is proportional to $i/\omega$ for $\omega\to 0$ then $T^{01}(0,t)$ tends to a non zero value as $t\to +\infty$. Starting from (\ref{Tff2}) we have
\ee{\fl T^{01}(0,t)= - m_{0} i \int{\frac{dk_1}{2\pi}\frac{dk_2}{2\pi} \frac{ \tilde \eta(k_1+k_2)}{2 (\omega_{01}+\omega_{02})} k_{2} \omega_{1} \left[ \left(\frac{1}{\omega_{01} \omega_{02}} + \frac{1}{\omega_1 \omega_2}\right) \sin(\omega_{1}+\omega_{2})t  + \right .} \nn \\
 \left . + \left(\frac{1}{\omega_{01} \omega_{02}} - \frac{1}{\omega_1 \omega_2}\right) \sin(\omega_{1}-\omega_{2})t \right]}
Its Fourier transform for small frequencies $\omega$ is 
\ee{\fl \tilde T^{01}(0;\omega) \sim - \frac{m_{0}}{8\pi} \int{{dk_1}{dk_2} \frac{ \tilde \eta(k_1+k_2)}{(\omega_{01}+\omega_{02})} k_{2} \omega_{1} \left(\frac{1}{\omega_{01} \omega_{02}} - \frac{1}{\omega_1 \omega_2}\right) \times }\nn \\  
\times \Bigl(\delta(\omega_{1}-\omega_{2}-\omega)-\delta(\omega_{1}-\omega_{2}+\omega)\Bigr)}
Notice that we have skipped the two $\delta$-functions of the form $\delta(\omega_{1}+\omega_{2}\pm \omega)$ that come from $\sin(\omega_{1}+\omega_{2})t$ since they vanish for $\omega\to 0$ as $\omega_{1}+\omega_{2} \geq 2 m >0$. Let us specialise in the step distribution 
\eq{
\eta(x)=
 \cases{ 
-a/2 & \mbox{ if $x<0$,}  \\
+a/2 & \mbox{ if $x>0$.}  
 } }
whose Fourier transform is $\tilde \eta(k) = ai/k$. If we symmetrise the integrand under $k_{1} \to -k_{1}$, $k_{2} \to -k_{2}$ and $k_{1}\leftrightarrow k_{2}$ we finally obtain
\eq{\fl \tilde T^{01}(0;\omega) \sim \frac{m_{0} a i}{8\pi} \int{{dk_1}{dk_2} \frac{k_{1}^{2}\omega_{2}+k_{2}^{2}\omega_{1}}{{ (\omega_{01}+\omega_{02})} (k_{1}^{2}-k_{2}^{2})} \left(\frac{1}{\omega_{01} \omega_{02}} - \frac{1}{\omega_1 \omega_2}\right) \delta(\omega_{1}-\omega_{2}-\omega)}}
Now the $\delta$-function prescribes that $\omega_{1}=\omega_{2}+\omega$ and substituting to the integrand we find to first order in $\omega$
\eq{\fl \tilde T^{01}(0;\omega\to 0) \sim - \frac{m_{0} a i (m_{0}^{2}-m^{2})}{16 \pi \omega}\int{{dk_1}{dk_2} \frac{(\omega_{1}^{2}-m^{2})}{\omega_{01}^{3} \omega_{1}^{2}} \delta(\omega_{1}-\omega_{2}-\omega)}}
This shows that $T^{01}(0;\omega)$ is of order $1/\omega$ for $\omega$ close to zero, which means that as before $T^{01}(0,t)$ does not tend to zero for large times. Evaluating the integrals carefully we end up with the simple relation
\eqq[Tmass]{\tilde T^{01}(0;\omega\to 0) \sim -\frac{ (m_{0}-m) a i}{4 \pi \omega}}
For $m=0$ we recover the already found result (\ref{Tomega}). Notice that the expression above gives zero if $m=m_{0}$.

\section{Comparison of the CFT and FFT results for the energy density after a quantum quench}\label{App4}

We will calculate the energy density after a homogeneous quench of the mass from $m_0$ to zero, using the CFT and FFT methods. The CFT result is obtained by the expectation value of the $T^{00}$ component of the stress-energy tensor on a strip of width $2 \epsilon$ 
\eq{\langle T^{00}_{strip} \rangle = -\frac{c \pi }{24 (2\epsilon)^2}}
The analytic continuation from imaginary to real time reverses the sign and when we also set $c=1$ and take into account the correspondence relation $\epsilon =1/m_0$, we obtain
\eq{\langle T^{00} \rangle = \frac{\pi m_0^2}{96}}

In free field theory the energy density $\langle T^{00} \rangle$ can be obtained from the propagator in the same way as the energy flow $\langle T^{01} \rangle$, but here we will show a slightly different derivation. The energy of the system which is conserved, is obviously equal to the expectation value of the hamiltonian after the quench $H = \frac{1}{2} \int{(\partial \phi)^2 dx } $ with respect to the ground state $|\Psi_0\rangle$ of the hamiltonian before the quench $H_0 = \frac{1}{2} \int{((\partial \phi)^2+ m_0^2 \phi^2 )dx } $. If we decompose the latter as follows
\eq{\langle \Psi_0 | H |\Psi_0 \rangle = \langle \Psi_0 | (H-H_0) |\Psi_0 \rangle + \langle \Psi_0 | H_0 |\Psi_0 \rangle}
then the first part is
\eq{-\frac{1}{2} m_0^2 \int{ dx }\; \langle \Psi_0 | \phi^2 |\Psi_0 \rangle }
which is easy to calculate since
\eq{\langle \Psi_0 | \phi^2 |\Psi_0 \rangle = \int{\frac{d^2 k}{(2 \pi)^2} \frac{1}{k^2 + m_0^2} }}
is a loop of the Feynman propagator with mass $m_0$ in (1+1)-$d$, while the second part is the reduced free energy of a system with hamiltonian $H_0 = H + \frac{1}{2}m_0^2 \int{ \phi^2 dx}$ where the last term is considered as a perturbation over $H$. The reduced free energy per unit length in this case is 
\eq{f(m_0^2) =  \frac{1}{2} \int{\frac{d^2 k}{(2 \pi)^2} \left[\log{(k^2 + m_0^2)}-\log{k^2} \right]}}
Putting all these together we find that the energy density is equal to
\eq{\frac{1}{2} \int{\frac{d^2 k}{(2 \pi)^2} \left[\log{(1 + m_0^2/k^2)} - \frac{m_0^2}{k^2+m_0^2}\right]}}
The last integral is both ultraviolet and infrared convergent and if we set $k^2=u m_0^2$ and integrate by parts, it gives
\eq{\frac{m_0^2}{8 \pi} \int\limits_0^\infty{\frac{du}{(1+u)^2}} = \frac{m_0^2}{8 \pi}}

Comparing the two results we notice that they differ by a factor $\pi^2/12$, exactly as for the energy flow.

\section*{References}
\bibliographystyle{unsrt}
\bibliography{bib}{}

\end{document}